# Statistical investigation of relationship between spread of coronavirus disease (COVID-19) and environmental factors based on study of four mostly affected places of China and five mostly affected places of Italy

Soumyabrata Bhattacharjee

s.bhattacharjee@rgi.edu.in

The Assam Royal Global University, Guwahati – 35, Assam, India

**Abstract:** COVID-19 is a new type of coronavirus disease which is caused by severe acute respiratory syndrome coronavirus 2 (SARS-CoV-2). It originated in China in the month of December 2019 and quickly started to spread within the country. On 31st December 2019, it was first reported to country office of World Health Organization (WHO) in China. Since then, it has spread to most of the countries around the globe. However, there has been a recent rise in trend in believing that it would go away during summer days, which has not yet been properly investigated. In this paper, relationship of daily number of confirmed cases of COVID-19 with three environmental factors, viz. maximum relative humidity ($RH_{max}$), maximum temperature ($T_{max}$) and highest wind speed ($WS_{max}$), considering the incubation period, have been investigated statistically, for four of the most affected places of China, viz. Beijing, Chongqing, Shanghai, Wuhan and five of the most affected places of Italy, viz. Bergamo, Cremona, Lodi, Milano. It has been found that the relationship with maximum relative humidity and highest wind is mostly negligible, whereas relationship with maximum temperature is ranging between negligible to moderate.

**Introduction:** Over the years, many different viruses of coronavirus family have surfaced and disrupted life of a lot of people. In 2002, a new disease called Severe Acute Respiratory Syndrome (SARS) started from Guangdong province of China. In 2003 it was found that, the disease is caused by SARS coronavirus (SARS-CoV) and it quickly became an epidemic, affecting life of more than 8000 people in 26 different countries [1]. In 2012, another disease called called Middle East Respiratory Syndrome (MERS), caused by MERS coronavirus (MERS-CoV), started from Saudi Arabia, became an epidemic and spread to 27 countries [2]. In the later part of 2019, an outbreak of pneumonia of unknown cause started appearing in Wuhan, China and was reported to the Country Office of WHO in China on 31st December 2019. On 30th January 2020, it was declared as Public Health Emergency of International Concern. On 11th February 2020, WHO named the disease as coronavirus disease (COVID-19) and the virus causing the disease as severe acute respiratory syndrome coronavirus 2 (SARS-CoV-2) [3]. On 11th March 2020, WHO declared COVID-19 as pandemic [4] and by 16th March 2020, the virus has spread to 156 countries/regions [5].

From 19th January 2020 onwards, there has been hearsay about relationship between coronavirus and summer [6]. However, there has been a sharp rise in interest shown by people all around the world, in the relationship, as shown by Google Trend [6], after President of the United States of America, who is currently one of the most influential person in the world [7], tweeted that the virus might be gone with warmer weather [8]. If such claims are not properly investigated then it might end up being rumour and ultimately hinder the disease control process [9], which has not yet been done widely.

One of the terms to describe severity of any infectious disease is effective reproduction number (R) [10]. Wang et al., tried to develop an equation to estimate the value of R with respect to the value of Temperature (T) and relative humidity (RH) [11], but they developed the equation based on data only from Chinese cities and applied that for places outside China. However, the environment and their effect are different at different places of the globe and the effect of incubation period has also not been considered while developing the equation. Incubation period of COVID-19 has been investigated by Baum et al. and it has been found that the median incubation period is approximately 5 days [12]. In the current paper, the relationship of daily number of confirmed COVID-19 cases, which has been assumed to be reflecting how contagious the disease is, from four of the most affected places of China and five of the most affected places of Italy has been investigated with $RH_{max}$, $T_{max}$ and $WS_{max}$. The environmental factors, that have been considered, are that of 5 days back from the date of reporting the case, to consider the effect of incubation period.

**Methodology:** Since 21st January, WHO has been publishing daily situation report on COVID-19 [13]. The data for the Chinese places has been collected from these reports and has been summarised in Table 1. All the weather data, reported in Table 1, have been collected from Weather Underground [14]

**Table 1:** Details of considered places from China

| Date | Place | Number of additional confirmed cases in last 24 hours | $RH_{max}$ (%) | $T_{max}$ (°F) | $WS_{max}$ (mph) |
|---|---|---|---|---|---|
| **20-01-2020** | Beijing | 3 | 65 | 35 | 4 |
| **21-01-2020** | Beijing | 0 | 74 | 36 | 4 |
| **22-01-2020** | Beijing | 5 | 75 | 37 | 4 |
| **02-02-2020** | Beijing | 27 | 75 | 41 | 4 |
| **03-02-2020** | Beijing | 29 | 86 | 46 | 7 |
| **04-02-2020** | Beijing | 16 | 80 | 43 | 9 |
| **05-02-2020** | Beijing | 25 | 74 | 44 | 4 |
| **06-02-2020** | Beijing | 21 | 33 | 40 | 7 |
| **07-02-2020** | Beijing | 23 | 89 | 34 | 7 |
| **08-02-2020** | Beijing | 18 | 49 | 36 | 11 |
| **09-02-2020** | Beijing | 11 | 61 | 34 | 13 |
| **10-02-2020** | Beijing | 11 | 83 | 24 | 7 |
| **11-02-2020** | Beijing | 5 | 87 | 24 | 7 |
| **12-02-2020** | Beijing | 10 | 83 | 35 | 2 |
| **13-02-2020** | Beijing | 14 | 89 | 44 | 4 |
| **14-02-2020** | Beijing | 6 | 82 | 42 | 7 |
| **15-02-2020** | Beijing | 3 | 90 | 52 | 2 |
| **16-02-2020** | Beijing | 5 | 87 | 49 | 4 |
| **17-02-2020** | Beijing | 1 | 90 | 51 | 2 |
| **18-02-2020** | Beijing | 6 | 93 | 44 | 7 |
| **19-02-2020** | Beijing | 6 | 94 | 44 | 18 |
| **20-02-2020** | Beijing | 2 | 37 | 31 | 13 |
| **21-02-2020** | Beijing | 1 | 33 | 38 | 13 |
| **22-02-2020** | Beijing | 3 | 52 | 41 | 11 |
| **23-02-2020** | Beijing | 0 | 74 | 49 | 7 |
| **24-02-2020** | Beijing | 0 | 83 | 46 | 7 |
| **25-02-2020** | Beijing | 1 | 80 | 40 | 4 |
| **26-02-2020** | Beijing | 0 | 93 | 47 | 7 |
| **27-02-2020** | Beijing | 10 | 29 | 48 | 9 |
| **28-02-2020** | Beijing | 0 | 55 | 48 | 4 |
| **29-02-2020** | Beijing | 1 | 71 | 48 | 7 |
| **01-03-2020** | Beijing | 2 | 47 | 48 | 4 |
| **02-03-2020** | Beijing | 1 | 5 | 46 | 9 |
| **03-03-2020** | Beijing | 0 | 51 | 41 | 7 |
| **04-03-2020** | Beijing | 3 | 72 | 45 | 4 |
| **05-03-2020** | Beijing | 1 | 95 | 38 | 4 |
| **06-03-2020** | Beijing | 4 | 96 | 50 | 11 |
| **07-03-2020** | Beijing | 4 | 67 | 47 | 9 |
| **08-03-2020** | Beijing | 2 | 91 | 45 | 13 |
| **09-03-2020** | Beijing | 0 | 34 | 43 | 9 |
| **10-03-2020** | Beijing | 1 | 44 | 47 | 11 |
| **11-03-2020** | Beijing | 6 | 72 | 50 | 7 |
| **12-03-2020** | Beijing | 0 | 89 | 58 | 7 |
| **13-03-2020** | Beijing | 1 | 94 | 42 | 4 |
| **14-03-2020** | Beijing | 1 | 94 | 52 | 7 |
| **20-01-2020** | Chongqing | 0 | 81 | 48 | 11 |
| **21-01-2020** | Chongqing | 1 | 87 | 48 | 4 |
| **22-01-2020** | Chongqing | 4 | 93 | 52 | 11 |
| **02-02-2020** | Chongqing | 24 | 81 | 54 | 7 |
| **03-02-2020** | Chongqing | 38 | 93 | 57 | 9 |
| **04-02-2020** | Chongqing | 37 | 87 | 52 | 9 |
| **05-02-2020** | Chongqing | 29 | 93 | 50 | 9 |
| **06-02-2020** | Chongqing | 23 | 87 | 48 | 9 |
| **07-02-2020** | Chongqing | 22 | 93 | 48 | 7 |
| **08-02-2020** | Chongqing | 15 | 93 | 50 | 7 |
| **09-02-2020** | Chongqing | 20 | 93 | 54 | 9 |
| **10-02-2020** | Chongqing | 22 | 87 | 54 | 11 |
| **11-02-2020** | Chongqing | 18 | 100 | 50 | 9 |
| **12-02-2020** | Chongqing | 19 | 100 | 54 | 9 |
| **13-02-2020** | Chongqing | 13 | 94 | 55 | 4 |
| **14-02-2020** | Chongqing | 11 | 100 | 55 | 7 |
| **15-02-2020** | Chongqing | 8 | 94 | 55 | 7 |
| **16-02-2020** | Chongqing | 7 | 87 | 59 | 7 |
| **17-02-2020** | Chongqing | 7 | 93 | 66 | 9 |
| **18-02-2020** | Chongqing | 2 | 76 | 66 | 9 |
| **19-02-2020** | Chongqing | 2 | 82 | 64 | 9 |
| **20-02-2020** | Chongqing | 5 | 93 | 54 | 11 |
| **21-02-2020** | Chongqing | 7 | 93 | 54 | 7 |
| **22-02-2020** | Chongqing | 5 | 81 | 55 | 7 |
| **23-02-2020** | Chongqing | 1 | 71 | 52 | 9 |
| **24-02-2020** | Chongqing | 2 | 93 | 55 | 9 |
| **25-02-2020** | Chongqing | 1 | 93 | 50 | 11 |
| **26-02-2020** | Chongqing | 0 | 93 | 54 | 9 |
| **27-02-2020** | Chongqing | 0 | 87 | 55 | 9 |
| **28-02-2020** | Chongqing | 0 | 100 | 52 | 9 |
| **29-02-2020** | Chongqing | 0 | 100 | 55 | 7 |
| **01-03-2020** | Chongqing | 0 | 94 | 59 | 9 |
| **02-03-2020** | Chongqing | 0 | 94 | 59 | 13 |
| **03-03-2020** | Chongqing | 0 | 94 | 57 | 7 |
| **04-03-2020** | Chongqing | 0 | 94 | 64 | 7 |
| **05-03-2020** | Chongqing | 0 | 100 | 64 | 9 |
| **06-03-2020** | Chongqing | 0 | 100 | 66 | 7 |
| **07-03-2020** | Chongqing | 0 | 94 | 61 | 16 |
| **08-03-2020** | Chongqing | 0 | 100 | 52 | 11 |
| **09-03-2020** | Chongqing | 0 | 94 | 55 | 7 |
| **10-03-2020** | Chongqing | 0 | 93 | 57 | 13 |
| **11-03-2020** | Chongqing | 0 | 82 | 63 | 9 |
| **12-03-2020** | Chongqing | 0 | 94 | 68 | 13 |
| **13-03-2020** | Chongqing | 0 | 77 | 61 | 13 |
| **14-03-2020** | Chongqing | 0 | 82 | 63 | 11 |
| **20-01-2020** | Shanghai | 1 | 93 | 48 | 11 |
| **21-01-2020** | Shanghai | 1 | 93 | 45 | 13 |
| **22-01-2020** | Shanghai | 7 | 93 | 45 | 13 |
| **02-02-2020** | Shanghai | 24 | 87 | 45 | 13 |
| **03-02-2020** | Shanghai | 16 | 76 | 50 | 20 |
| **04-02-2020** | Shanghai | 15 | 75 | 46 | 18 |
| **05-02-2020** | Shanghai | 25 | 80 | 50 | 13 |
| **06-02-2020** | Shanghai | 21 | 93 | 57 | 9 |
| **07-02-2020** | Shanghai | 15 | 81 | 57 | 11 |
| **08-02-2020** | Shanghai | 12 | 93 | 52 | 16 |
| **09-02-2020** | Shanghai | 11 | 87 | 57 | 11 |
| **10-02-2020** | Shanghai | 3 | 93 | 52 | 16 |
| **11-02-2020** | Shanghai | 7 | 93 | 45 | 16 |
| **12-02-2020** | Shanghai | 4 | 93 | 48 | 11 |
| **13-02-2020** | Shanghai | 7 | 81 | 50 | 11 |
| **14-02-2020** | Shanghai | 5 | 93 | 54 | 11 |
| **15-02-2020** | Shanghai | 8 | 93 | 63 | 13 |
| **16-02-2020** | Shanghai | 2 | 100 | 54 | 16 |
| **17-02-2020** | Shanghai | 3 | 100 | 63 | 9 |
| **18-02-2020** | Shanghai | 2 | 100 | 61 | 16 |
| **19-02-2020** | Shanghai | 0 | 100 | 64 | 13 |
| **20-02-2020** | Shanghai | 0 | 100 | 57 | 22 |
| **21-02-2020** | Shanghai | 1 | 93 | 43 | 22 |
| **22-02-2020** | Shanghai | 0 | 60 | 50 | 22 |
| **23-02-2020** | Shanghai | 1 | 65 | 54 | 11 |
| **24-02-2020** | Shanghai | 0 | 81 | 55 | 13 |
| **25-02-2020** | Shanghai | 0 | 87 | 61 | 11 |
| **26-02-2020** | Shanghai | 1 | 93 | 64 | 16 |
| **27-02-2020** | Shanghai | 1 | 82 | 64 | 16 |
| **28-02-2020** | Shanghai | 0 | 71 | 59 | 16 |
| **29-02-2020** | Shanghai | 0 | 82 | 72 | 13 |
| **01-03-2020** | Shanghai | 0 | 88 | 79 | 18 |
| **02-03-2020** | Shanghai | 0 | 87 | 52 | 16 |
| **03-03-2020** | Shanghai | 1 | 76 | 57 | 13 |
| **04-03-2020** | Shanghai | 0 | 94 | 55 | 18 |
| **05-03-2020** | Shanghai | 0 | 100 | 54 | 9 |
| **06-03-2020** | Shanghai | 1 | 94 | 52 | 16 |
| **07-03-2020** | Shanghai | 3 | 87 | 52 | 13 |
| **08-03-2020** | Shanghai | 0 | 76 | 55 | 9 |
| **09-03-2020** | Shanghai | 0 | 71 | 54 | 18 |
| **10-03-2020** | Shanghai | 0 | 65 | 54 | 11 |
| **11-03-2020** | Shanghai | 2 | 88 | 59 | 16 |
| **12-03-2020** | Shanghai | 0 | 94 | 63 | 13 |
| **13-03-2020** | Shanghai | 2 | 93 | 63 | 20 |
| **14-03-2020** | Shanghai | 4 | 100 | 57 | 16 |
| **20-01-2020** | Wuhan | 60 | 100 | 39 | 11 |
| **21-01-2020** | Wuhan | 12 | 87 | 37 | 9 |
| **22-01-2020** | Wuhan | 105 | 93 | 41 | 4 |
| **26-01-2020** | Wuhan | 323 | 87 | 41 | 11 |
| **27-01-2020** | Wuhan | 371 | 100 | 45 | 7 |
| **28-01-2020** | Wuhan | 1291 | 100 | 48 | 9 |
| **29-01-2020** | Wuhan | 840 | 100 | 46 | 11 |
| **30-01-2020** | Wuhan | 1032 | 93 | 43 | 9 |
| **31-01-2020** | Wuhan | 1220 | 93 | 39 | 9 |
| **01-02-2020** | Wuhan | 1347 | 87 | 41 | 9 |
| **02-02-2020** | Wuhan | 1921 | 93 | 46 | 7 |
| **03-02-2020** | Wuhan | 2103 | 100 | 54 | 9 |
| **04-02-2020** | Wuhan | 2345 | 100 | 57 | 7 |
| **05-02-2020** | Wuhan | 3156 | 81 | 57 | 9 |
| **06-02-2020** | Wuhan | 2987 | 70 | 57 | 11 |
| **07-02-2020** | Wuhan | 2447 | 87 | 54 | 7 |
| **08-02-2020** | Wuhan | 2841 | 100 | 57 | 7 |
| **09-02-2020** | Wuhan | 2147 | 87 | 59 | 7 |
| **10-02-2020** | Wuhan | 2531 | 87 | 61 | 13 |
| **11-02-2020** | Wuhan | 2097 | 93 | 45 | 11 |
| **12-02-2020** | Wuhan | 1638 | 93 | 43 | 9 |
| **13-02-2020** | Wuhan | 1508 | 87 | 48 | 7 |
| **14-02-2020** | Wuhan | 4823 | 100 | 57 | 7 |
| **15-02-2020** | Wuhan | 2420 | 93 | 52 | 7 |
| **16-02-2020** | Wuhan | 1843 | 93 | 54 | 7 |
| **17-02-2020** | Wuhan | 1933 | 94 | 57 | 7 |
| **18-02-2020** | Wuhan | 1807 | 94 | 64 | 13 |
| **19-02-2020** | Wuhan | 1693 | 100 | 61 | 20 |
| **20-02-2020** | Wuhan | 349 | 100 | 46 | 20 |
| **21-02-2020** | Wuhan | 631 | 93 | 46 | 7 |
| **22-02-2020** | Wuhan | 366 | 93 | 54 | 7 |
| **23-02-2020** | Wuhan | 630 | 87 | 57 | 9 |
| **24-02-2020** | Wuhan | 398 | 66 | 59 | 7 |
| **25-02-2020** | Wuhan | 499 | 81 | 64 | 11 |
| **26-02-2020** | Wuhan | 401 | 87 | 59 | 9 |
| **27-02-2020** | Wuhan | 409 | 100 | 63 | 9 |
| **28-02-2020** | Wuhan | 318 | 67 | 68 | 9 |
| **29-02-2020** | Wuhan | 423 | 73 | 75 | 13 |
| **01-03-2020** | Wuhan | 570 | 94 | 72 | 13 |
| **02-03-2020** | Wuhan | 196 | 100 | 59 | 13 |
| **03-03-2020** | Wuhan | 114 | 100 | 52 | 11 |
| **04-03-2020** | Wuhan | 115 | 100 | 45 | 9 |
| **05-03-2020** | Wuhan | 134 | 100 | 54 | 7 |
| **06-03-2020** | Wuhan | 126 | 100 | 55 | 13 |
| **07-03-2020** | Wuhan | 74 | 87 | 50 | 11 |
| **08-03-2020** | Wuhan | 41 | 93 | 48 | 4 |
| **09-03-2020** | Wuhan | 36 | 76 | 59 | 7 |
| **10-03-2020** | Wuhan | 17 | 81 | 61 | 9 |
| **11-03-2020** | Wuhan | 13 | 88 | 57 | 9 |
| **12-03-2020** | Wuhan | 8 | 88 | 66 | 9 |
| **13-03-2020** | Wuhan | 5 | 100 | 64 | 13 |
| **14-03-2020** | Wuhan | 4 | 100 | 57 | 13 |

Similarly, the data for Italy has been collected from the official GitHub repository of Department of Civil Protection, Italy [15] and the summary is presented in Table 2 along with weather data which is obtained from Weather Underground [14].

**Table 2:** Details of considered places from Italy

| Date | Place | Number of additional confirmed cases in last 24 hours | $RH_{max}$ (%) | $T_{max}$ (°F) | $WS_{max}$ (mph) |
|---|---|---|---|---|---|
| **24-02-2020** | Bergamo | 0 | 93 | 57 | 17 |
| **25-02-2020** | Bergamo | 18 | 70 | 55 | 8 |
| **26-02-2020** | Bergamo | 2 | 75 | 55 | 7 |
| **27-02-2020** | Bergamo | 52 | 81 | 55 | 8 |
| **28-02-2020** | Bergamo | 31 | 87 | 61 | 14 |
| **29-02-2020** | Bergamo | 7 | 93 | 61 | 10 |
| **01-03-2020** | Bergamo | 99 | 93 | 57 | 9 |
| **02-03-2020** | Bergamo | 34 | 100 | 55 | 24 |
| **03-03-2020** | Bergamo | 129 | 57 | 52 | 9 |
| **04-03-2020** | Bergamo | 51 | 65 | 57 | 21 |
| **05-03-2020** | Bergamo | 114 | 87 | 52 | 8 |
| **06-03-2020** | Bergamo | 86 | 100 | 46 | 9 |
| **07-03-2020** | Bergamo | 138 | 100 | 48 | 13 |
| **08-03-2020** | Bergamo | 236 | 100 | 54 | 10 |
| **09-03-2020** | Bergamo | 248 | 87 | 54 | 8 |
| **10-03-2020** | Bergamo | 227 | 100 | 50 | 9 |
| **11-03-2020** | Bergamo | 343 | 100 | 48 | 8 |
| **12-03-2020** | Bergamo | 321 | 93 | 59 | 9 |
| **13-03-2020** | Bergamo | 232 | 76 | 55 | 8 |
| **24-02-2020** | Brescia | 0 | 93 | 57 | 17 |
| **25-02-2020** | Brescia | 0 | 70 | 55 | 8 |
| **26-02-2020** | Brescia | 2 | 75 | 55 | 7 |
| **27-02-2020** | Brescia | 8 | 81 | 55 | 8 |
| **28-02-2020** | Brescia | 3 | 87 | 61 | 14 |
| **29-02-2020** | Brescia | 1 | 100 | 59 | 20 |
| **01-03-2020** | Brescia | 35 | 93 | 55 | 7 |
| **02-03-2020** | Brescia | 11 | 100 | 59 | 25 |
| **03-03-2020** | Brescia | 26 | 69 | 54 | 15 |
| **04-03-2020** | Brescia | 41 | 65 | 61 | 18 |
| **05-03-2020** | Brescia | 28 | 71 | 52 | 9 |
| **06-03-2020** | Brescia | 27 | 100 | 46 | 9 |
| **07-03-2020** | Brescia | 231 | 100 | 46 | 24 |
| **08-03-2020** | Brescia | 88 | 100 | 48 | 15 |
| **09-03-2020** | Brescia | 238 | 93 | 55 | 9 |
| **10-03-2020** | Brescia | 51 | 93 | 52 | 15 |
| **11-03-2020** | Brescia | 561 | 100 | 50 | 16 |
| **12-03-2020** | Brescia | 247 | 100 | 59 | 9 |
| **13-03-2020** | Brescia | 186 | 93 | 57 | 10 |
| **24-02-2020** | Cremona | 0 | 100 | 57 | 17 |
| **25-02-2020** | Cremona | 53 | 70 | 57 | 9 |
| **26-02-2020** | Cremona | 4 | 81 | 59 | 8 |
| **27-02-2020** | Cremona | 34 | 81 | 57 | 7 |
| **28-02-2020** | Cremona | 32 | 93 | 61 | 18 |
| **29-02-2020** | Cremona | 13 | 100 | 59 | 14 |
| **01-03-2020** | Cremona | 78 | 100 | 54 | 6 |
| **02-03-2020** | Cremona | 9 | 100 | 61 | 29 |
| **03-03-2020** | Cremona | 64 | 64 | 54 | 12 |
| **04-03-2020** | Cremona | 46 | 61 | 59 | 18 |
| **05-03-2020** | Cremona | 73 | 66 | 54 | 8 |
| **06-03-2020** | Cremona | 46 | 100 | 46 | 8 |
| **07-03-2020** | Cremona | 110 | 100 | 52 | 20 |
| **08-03-2020** | Cremona | 103 | 100 | 50 | 10 |
| **09-03-2020** | Cremona | 251 | 93 | 55 | 8 |
| **10-03-2020** | Cremona | 41 | 93 | 52 | 16 |
| **11-03-2020** | Cremona | 104 | 93 | 52 | 17 |
| **12-03-2020** | Cremona | 241 | 93 | 59 | 12 |
| **13-03-2020** | Cremona | 42 | 81 | 57 | 8 |
| **24-02-2020** | Lodi | 0 | 100 | 57 | 8 |
| **25-02-2020** | Lodi | 125 | 76 | 57 | 7 |
| **26-02-2020** | Lodi | 3 | 93 | 61 | 9 |
| **27-02-2020** | Lodi | 31 | 93 | 57 | 7 |
| **28-02-2020** | Lodi | 23 | 87 | 59 | 13 |
| **29-02-2020** | Lodi | 55 | 82 | 66 | 13 |
| **01-03-2020** | Lodi | 107 | 93 | 57 | 5 |
| **02-03-2020** | Lodi | 40 | 100 | 57 | 20 |
| **03-03-2020** | Lodi | 98 | 65 | 54 | 9 |
| **04-03-2020** | Lodi | 77 | 65 | 61 | 14 |
| **05-03-2020** | Lodi | 99 | 93 | 50 | 9 |
| **06-03-2020** | Lodi | 81 | 100 | 46 | 6 |
| **07-03-2020** | Lodi | 72 | 100 | 48 | 18 |
| **08-03-2020** | Lodi | 42 | 100 | 54 | 13 |
| **09-03-2020** | Lodi | 75 | 100 | 55 | 10 |
| **10-03-2020** | Lodi | 35 | 93 | 50 | 10 |
| **11-03-2020** | Lodi | 72 | 93 | 52 | 9 |
| **12-03-2020** | Lodi | 88 | 100 | 61 | 12 |
| **13-03-2020** | Lodi | 10 | 100 | 57 | 6 |
| **24-02-2020** | Milano | 0 | 100 | 57 | 8 |
| **25-02-2020** | Milano | 8 | 76 | 57 | 7 |
| **26-02-2020** | Milano | 0 | 93 | 61 | 9 |
| **27-02-2020** | Milano | 7 | 93 | 57 | 7 |
| **28-02-2020** | Milano | 14 | 87 | 59 | 13 |
| **29-02-2020** | Milano | 1 | 82 | 66 | 13 |
| **01-03-2020** | Milano | 16 | 93 | 57 | 5 |
| **02-03-2020** | Milano | 12 | 100 | 57 | 20 |
| **03-03-2020** | Milano | 35 | 65 | 54 | 9 |
| **04-03-2020** | Milano | 52 | 65 | 61 | 14 |
| **05-03-2020** | Milano | 52 | 93 | 50 | 9 |
| **06-03-2020** | Milano | 70 | 100 | 46 | 6 |
| **07-03-2020** | Milano | 94 | 100 | 48 | 18 |
| **08-03-2020** | Milano | 45 | 100 | 54 | 13 |
| **09-03-2020** | Milano | 100 | 100 | 55 | 10 |
| **10-03-2020** | Milano | 86 | 93 | 50 | 10 |
| **11-03-2020** | Milano | 333 | 93 | 52 | 9 |
| **12-03-2020** | Milano | 221 | 100 | 61 | 12 |
| **13-03-2020** | Milano | 161 | 100 | 57 | 6 |

Once the data has been collected, then to find out how well the daily number of newly confirmed cases of COVID-19 relates to $RH_{max}$, $T_{max}$ and $WS_{max}$, at first the daily number of cases is plotted against each one of the considered environmental factor and visually inspected, then Pearson's correlation coefficient (r) for each such pair has been calculated [16] using MATLAB (R2019b). After that, the each of the values are interpreted according to the rule of thumb mentioned by Mukaka [17].

To further strengthen the finding, hypothesis test has also been done on them using 95% confidence interval. Null hypothesis, for each such pair, being that the environmental factors do not influence the spread of disease. But, while testing hypothesis, the results are not interpreted depending only on the p-values, as cautioned by Wasserstein et al. [18]. Instead Bayes Factor (BF), for each one of them is calculated using the expression $-\frac{1}{ep\ln(p)}$ [19] and then the relationship is once again interpreted based on the value of BF as mentioned in the classification scheme by Jamil et al. [20].

**Results and discussion:** From figure 1 – 27, it can be seen that the influence of environmental factors is neither that strong nor that can be outrightly rejected.

The value and interpretation of Pearson's correlation coefficient and Bayes Factor is listed in Table 3.

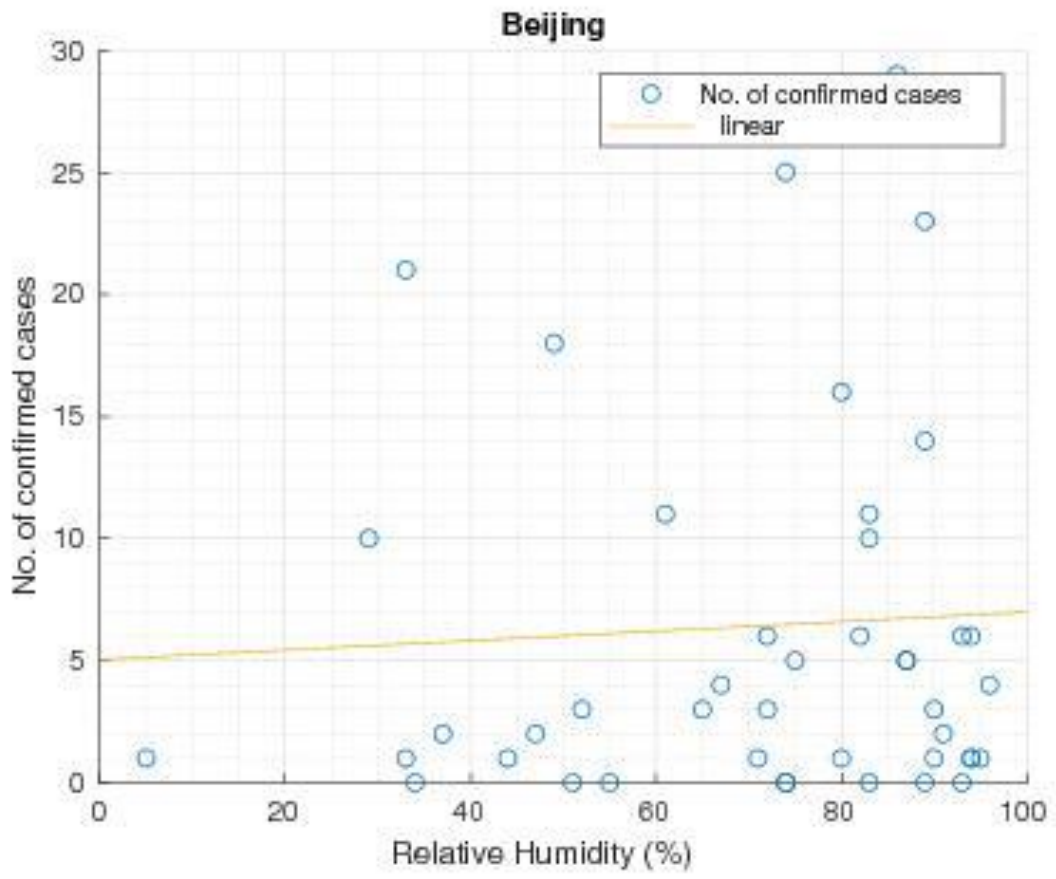

*Figure 1: Effect of maximum relative humidity in Beijing*

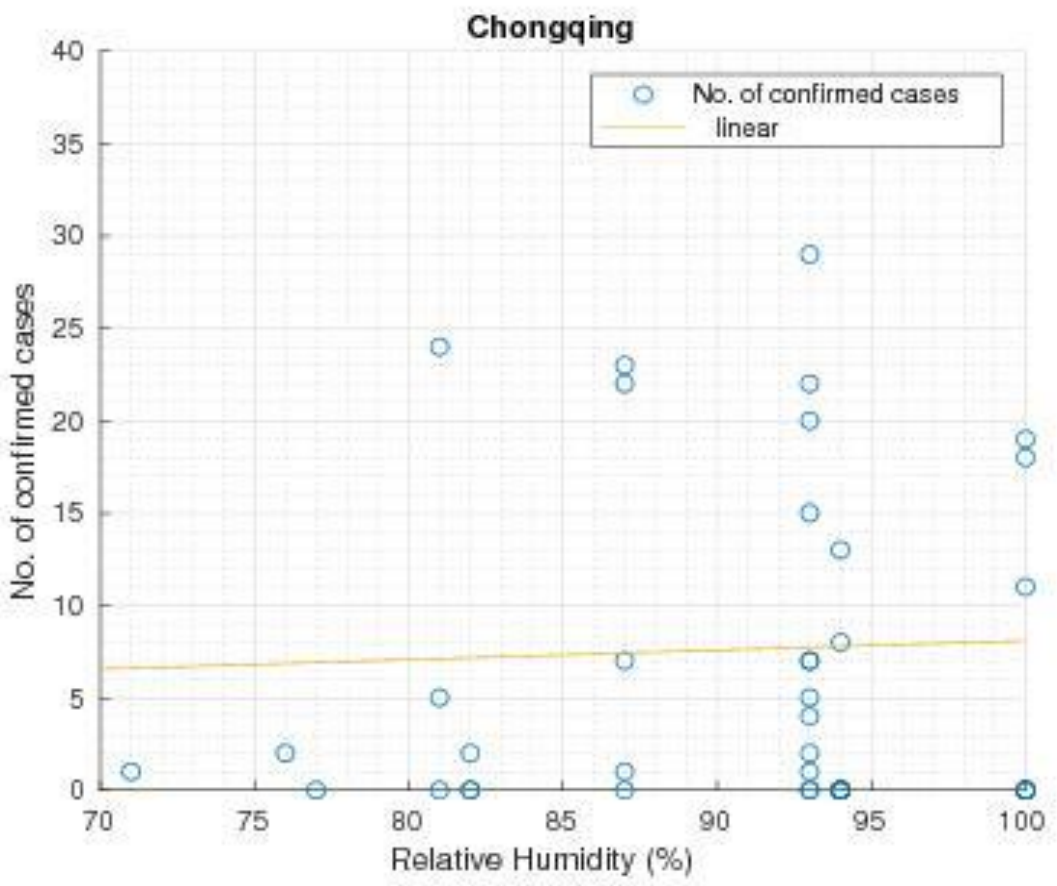

*Figure 2: Effect of maximum relative humidity in Chongqing*

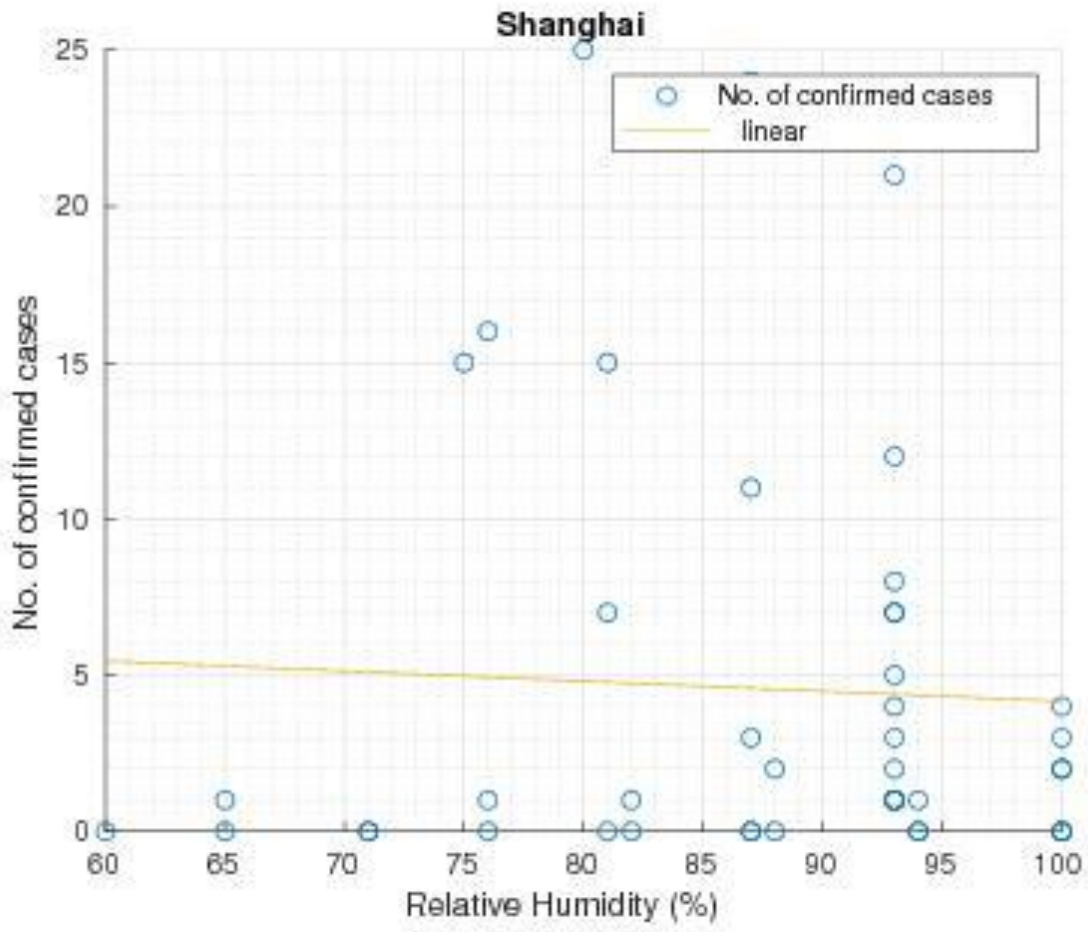

*Figure 3: Effect of maximum relative humidity in Shanghai*

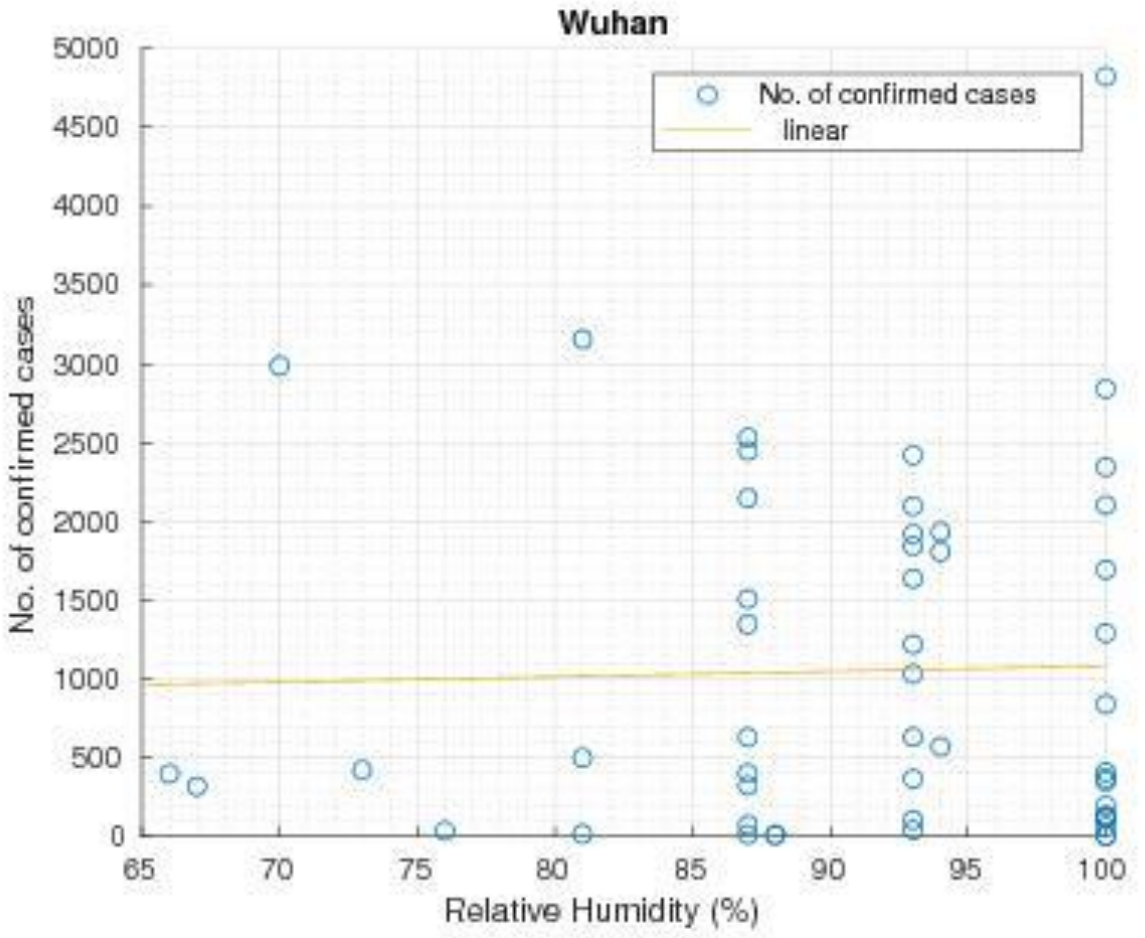

*Figure 4: Effect of maximum relative humidity in Wuhan*

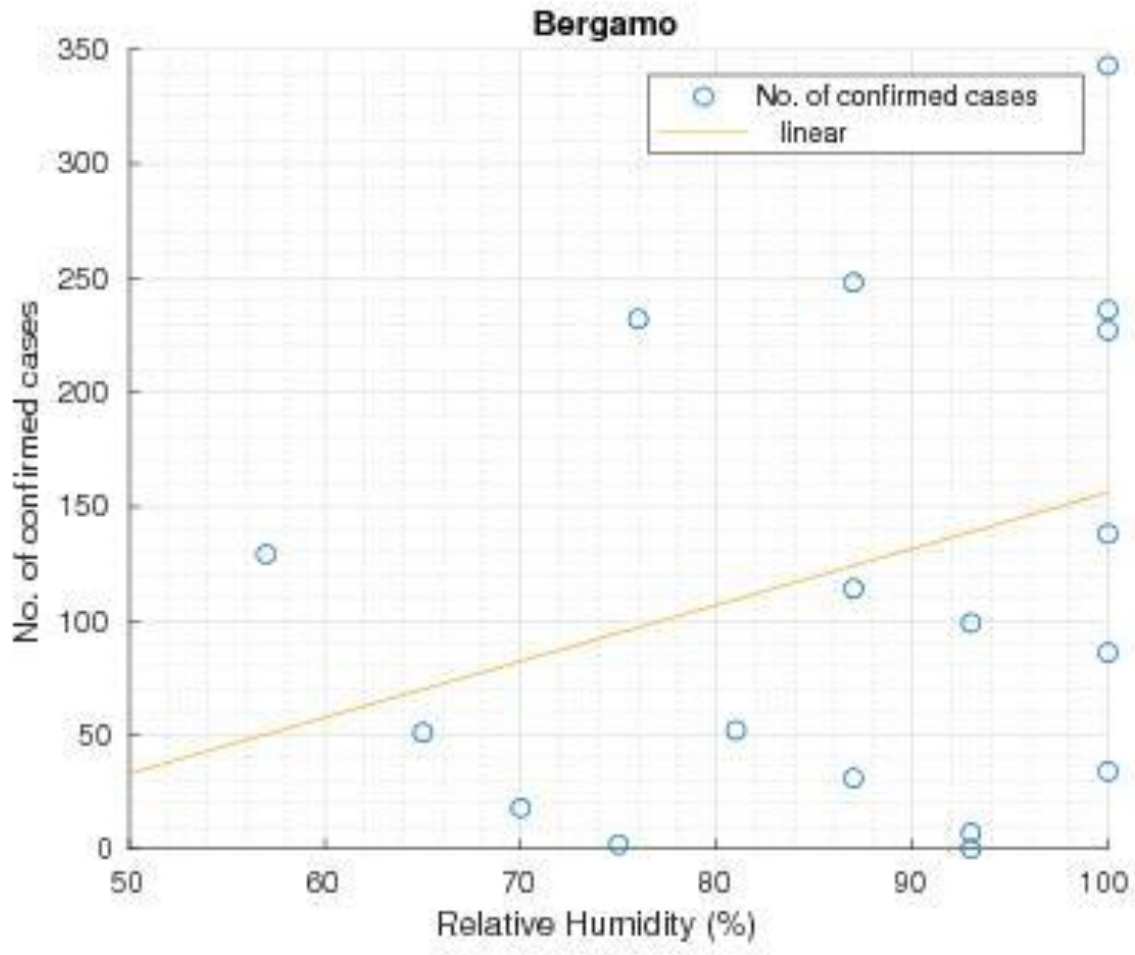

*Figure 5: Effect of maximum relative humidity in Bergamo*

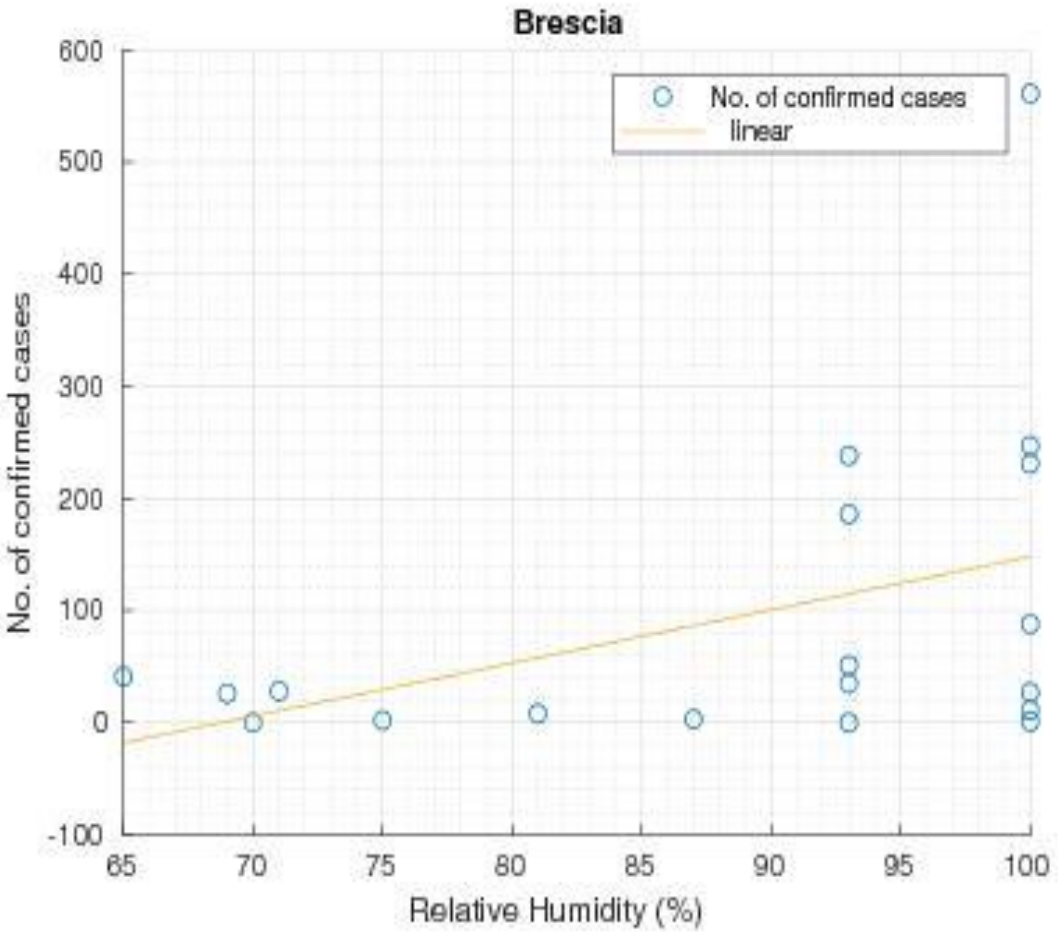

*Figure 6: Effect of maximum relative humidity in Brescia*

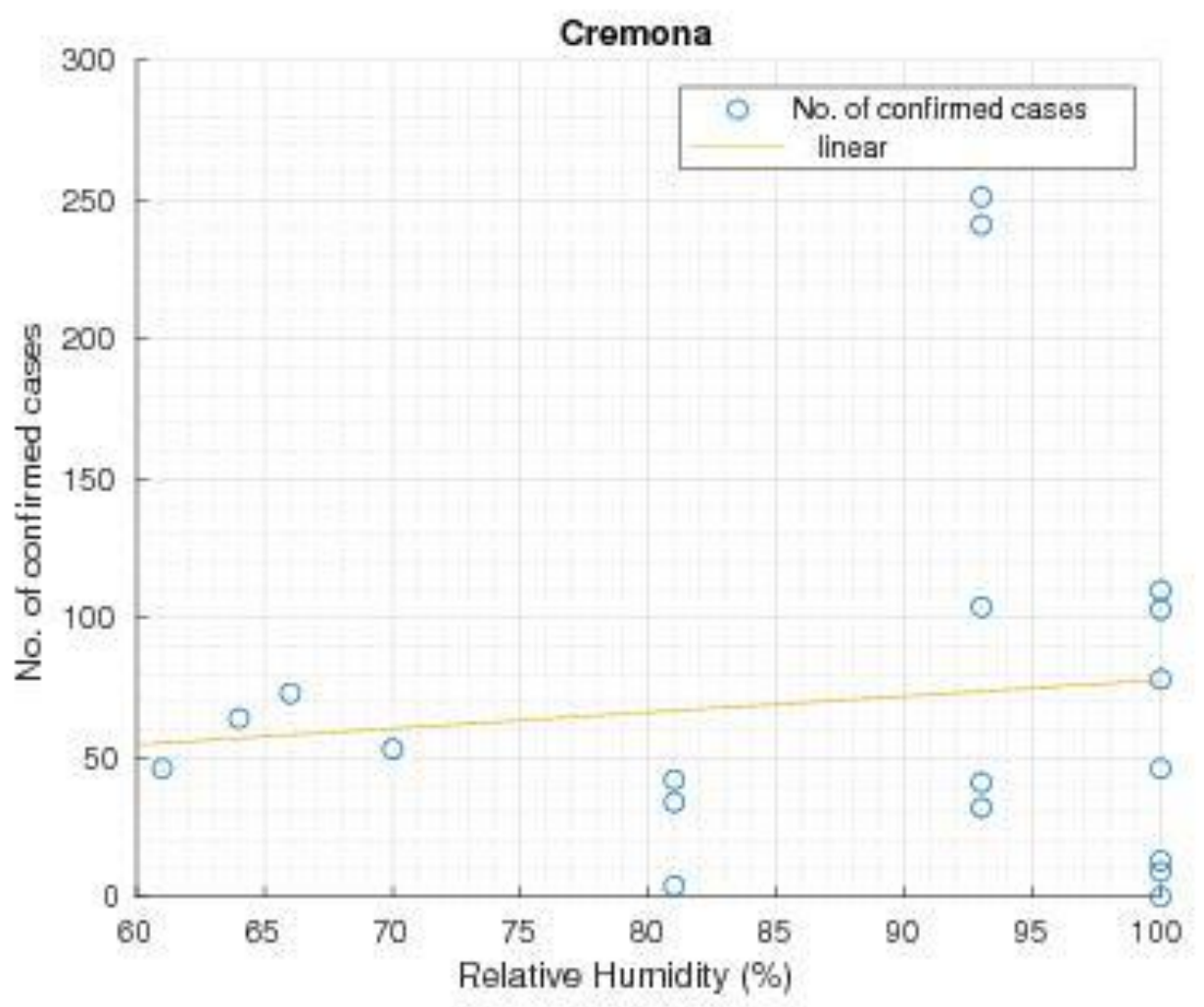

*Figure 7: Effect of maximum relative humidity in Cremona*

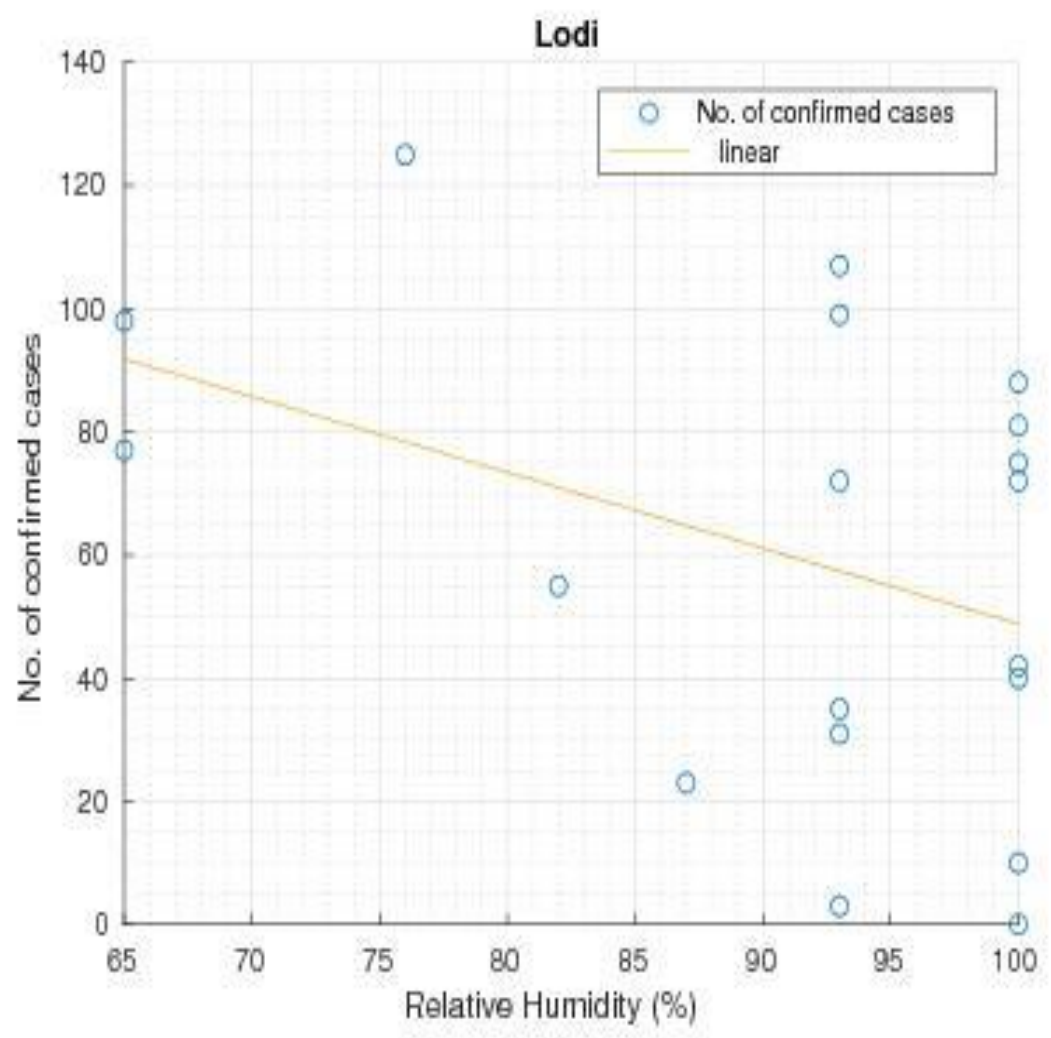

*Figure 8: Effect of maximum relative humidity in Lodi*

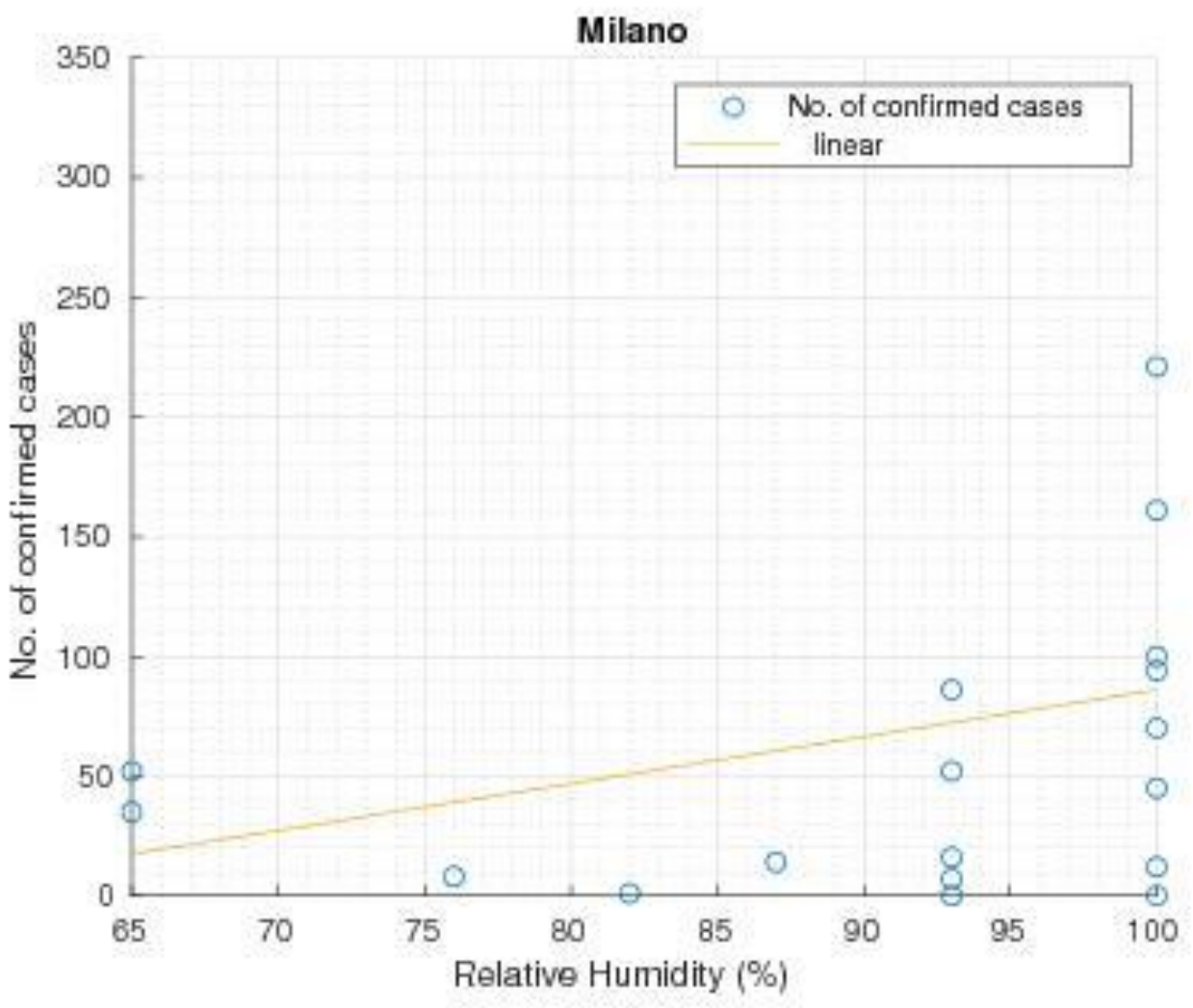

*Figure 9: Effect of maximum relative humidity in Milano*

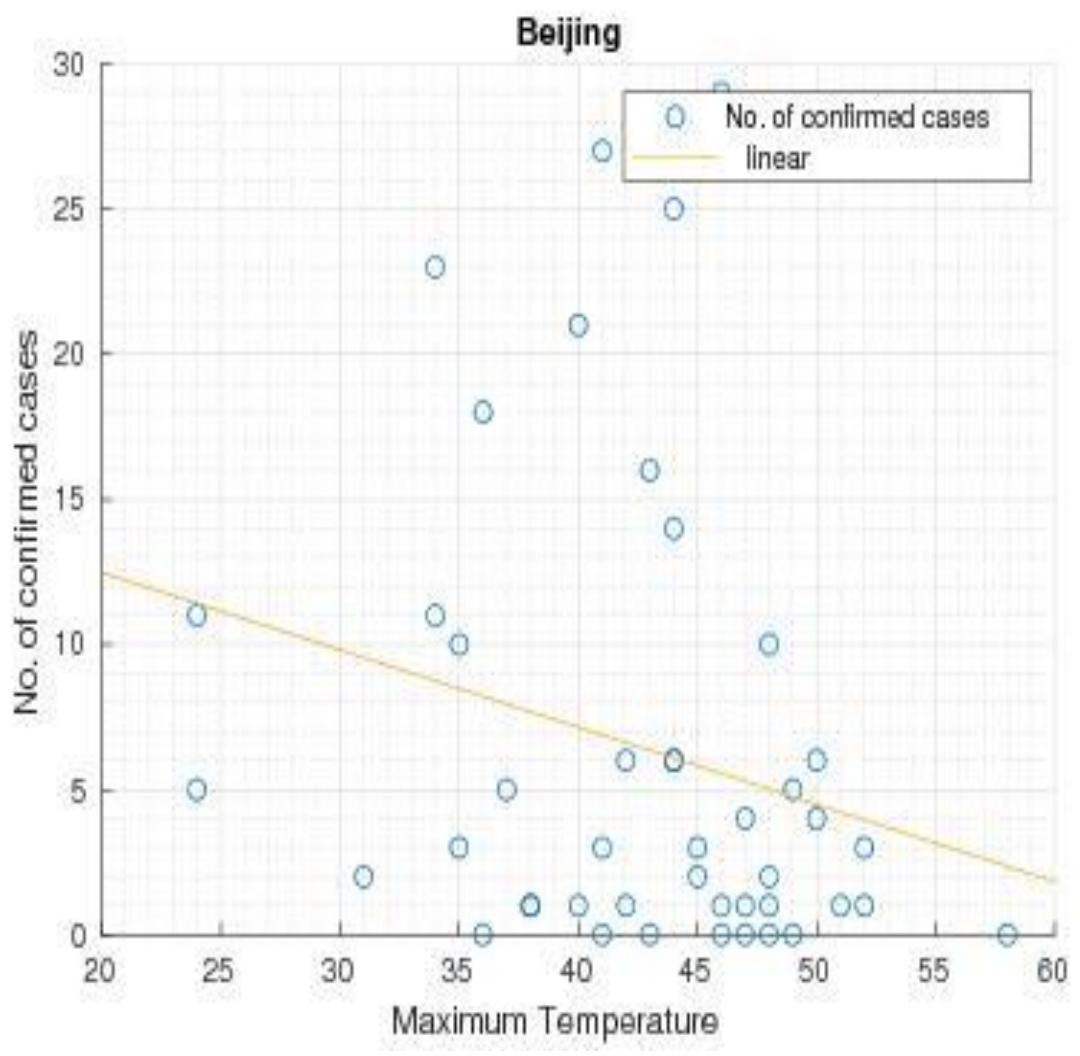

*Figure 10: Effect of maximum temperature in Beijing*

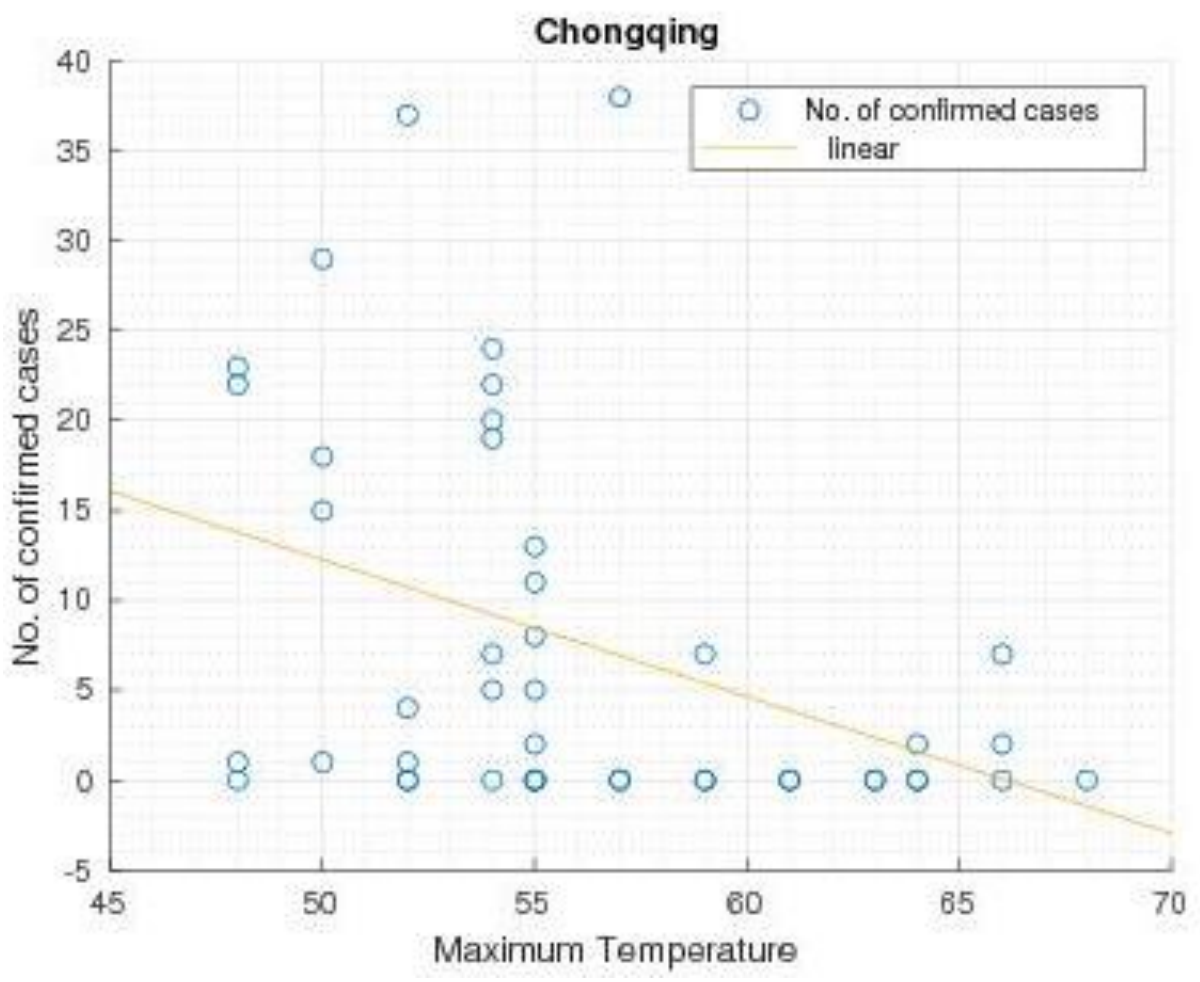

*Figure 11: Effect of maximum temperature in Chongqing*

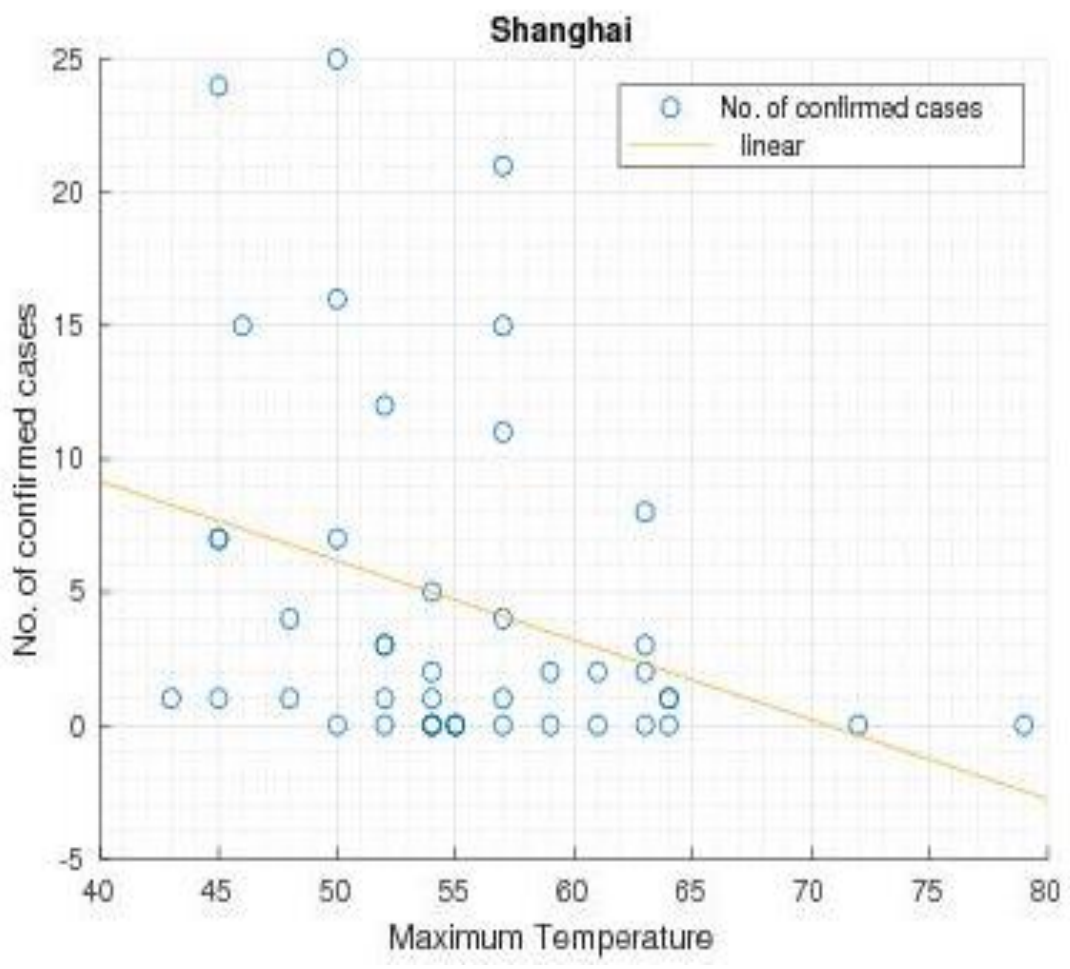

*Figure 12: Effect of maximum temperature in Shanghai*

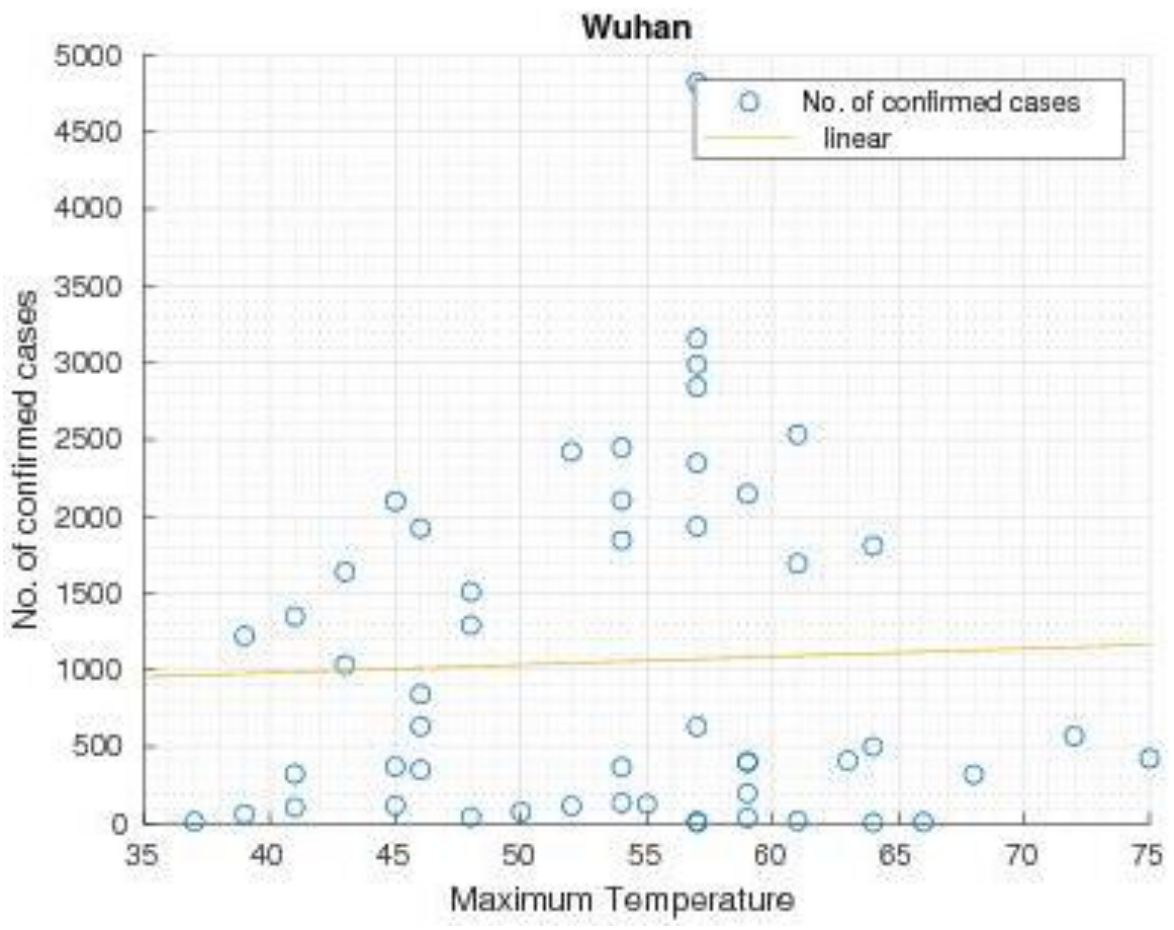

*Figure 13: Effect of maximum temperature in Wuhan*

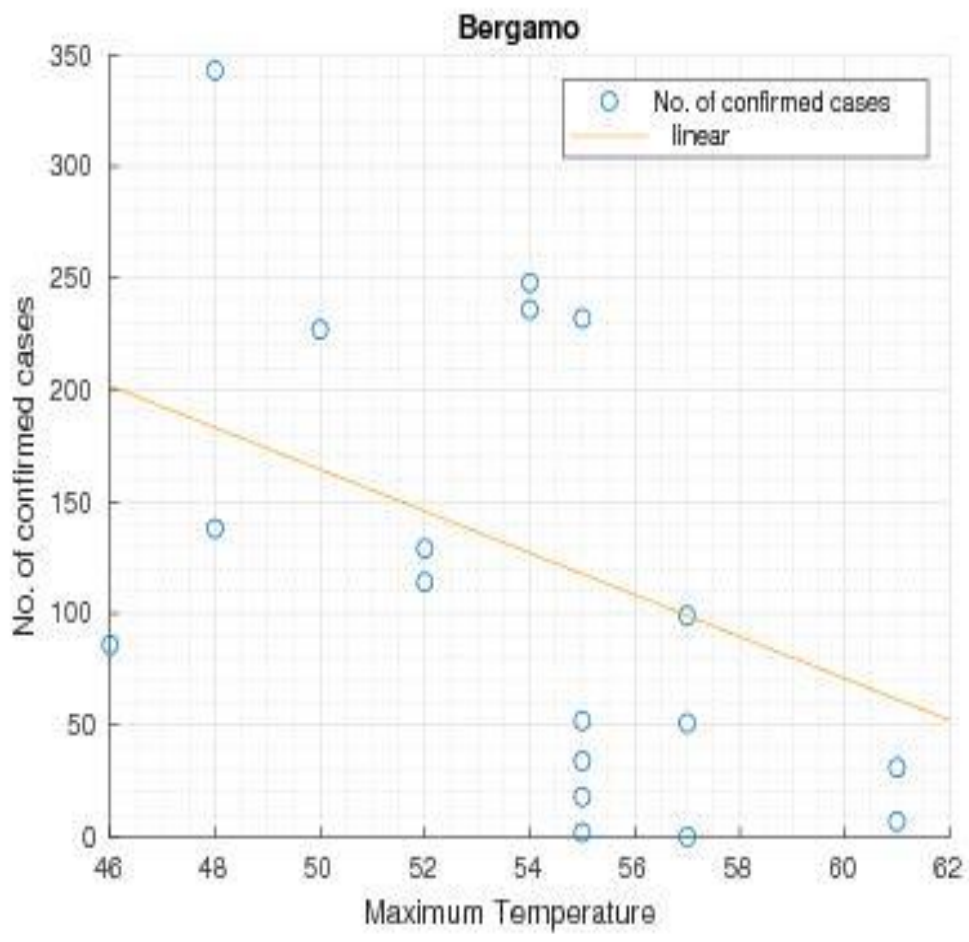

*Figure 14: Effect of maximum temperature in Bergamo*

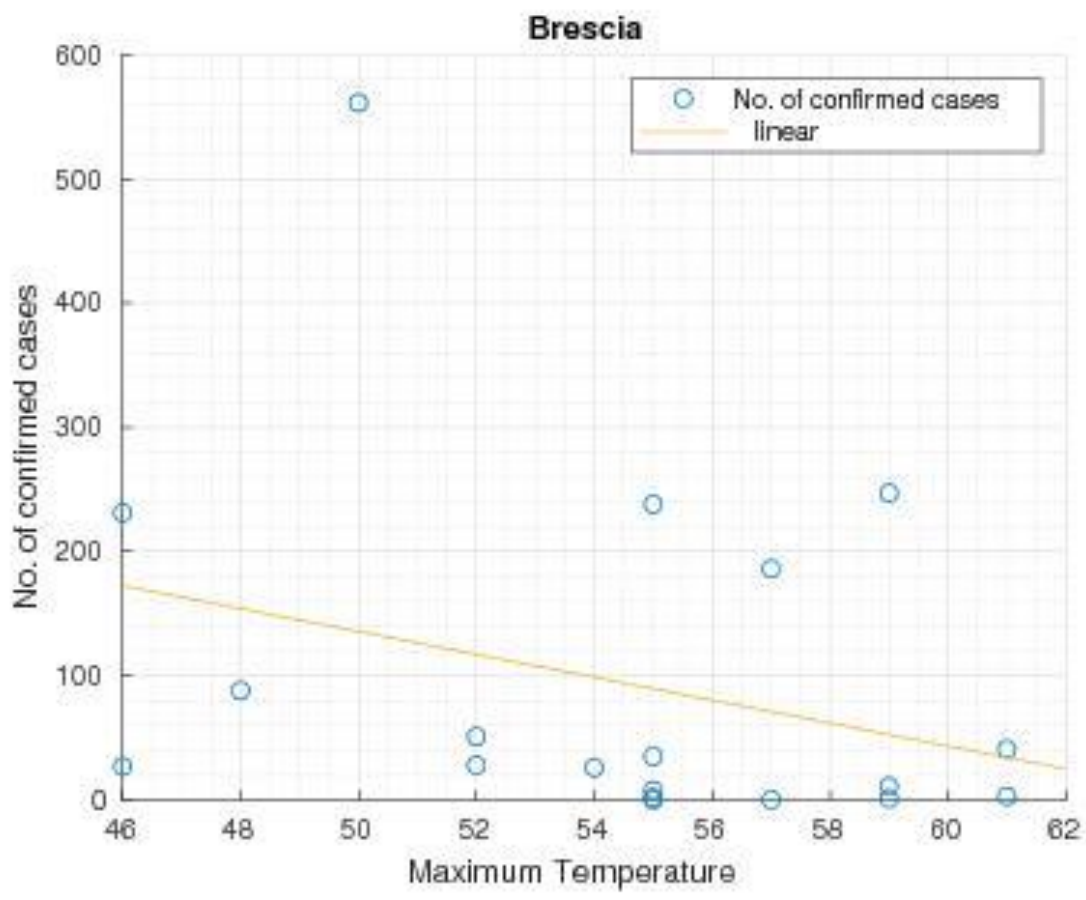

*Figure 15: Effect of maximum temperature in Brescia*

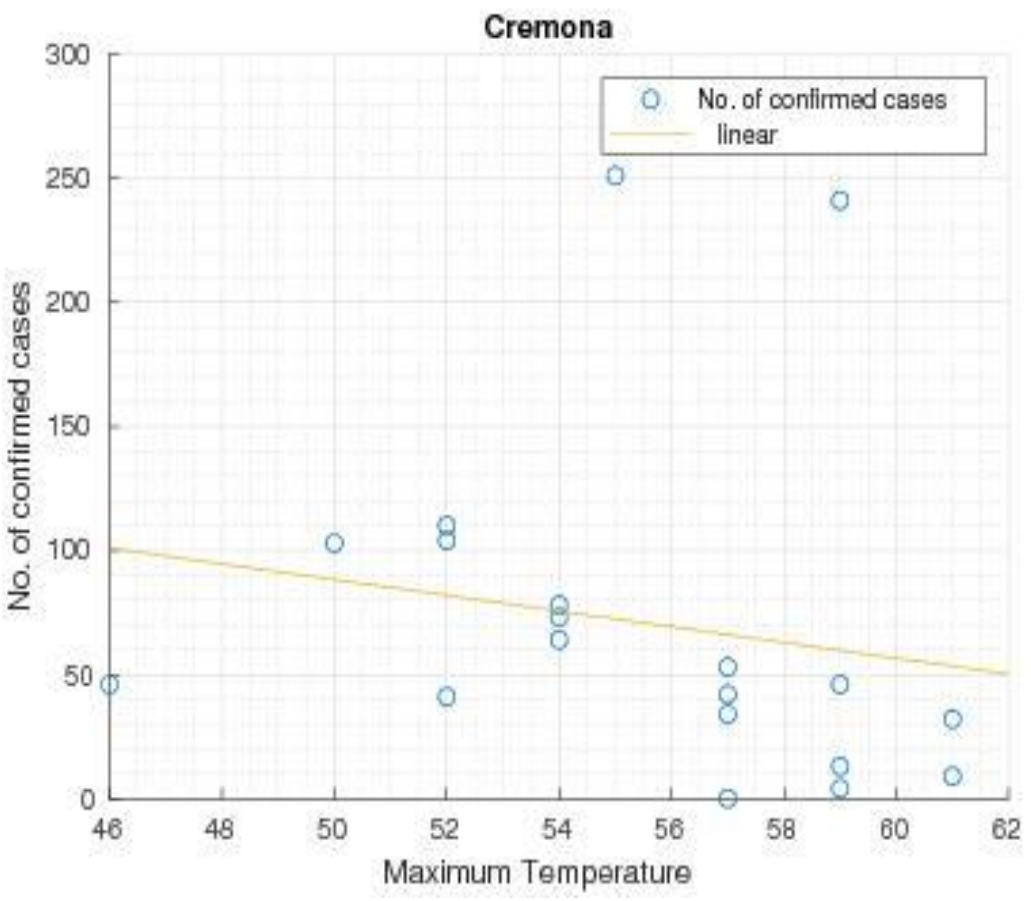

*Figure 16: Effect of maximum temperature in Cremona*

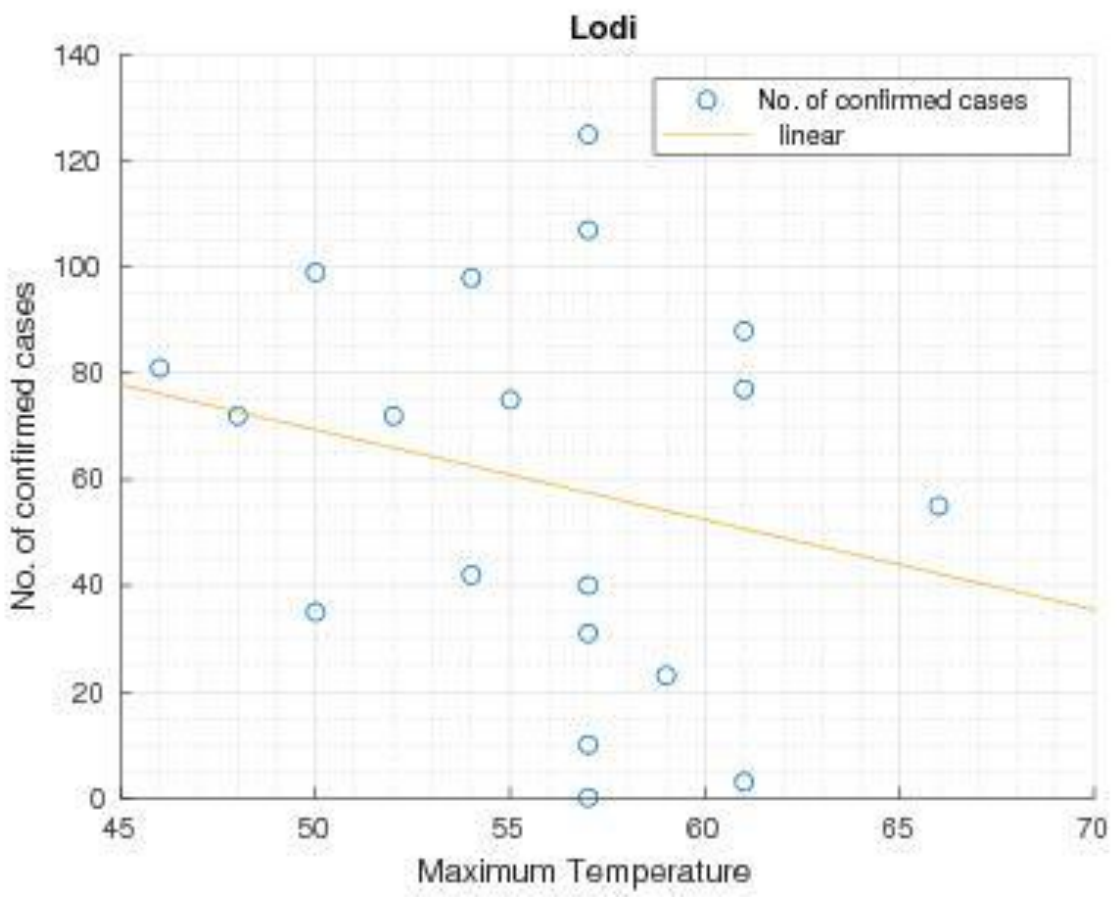

*Figure 17: Effect of maximum temperature in Lodi*

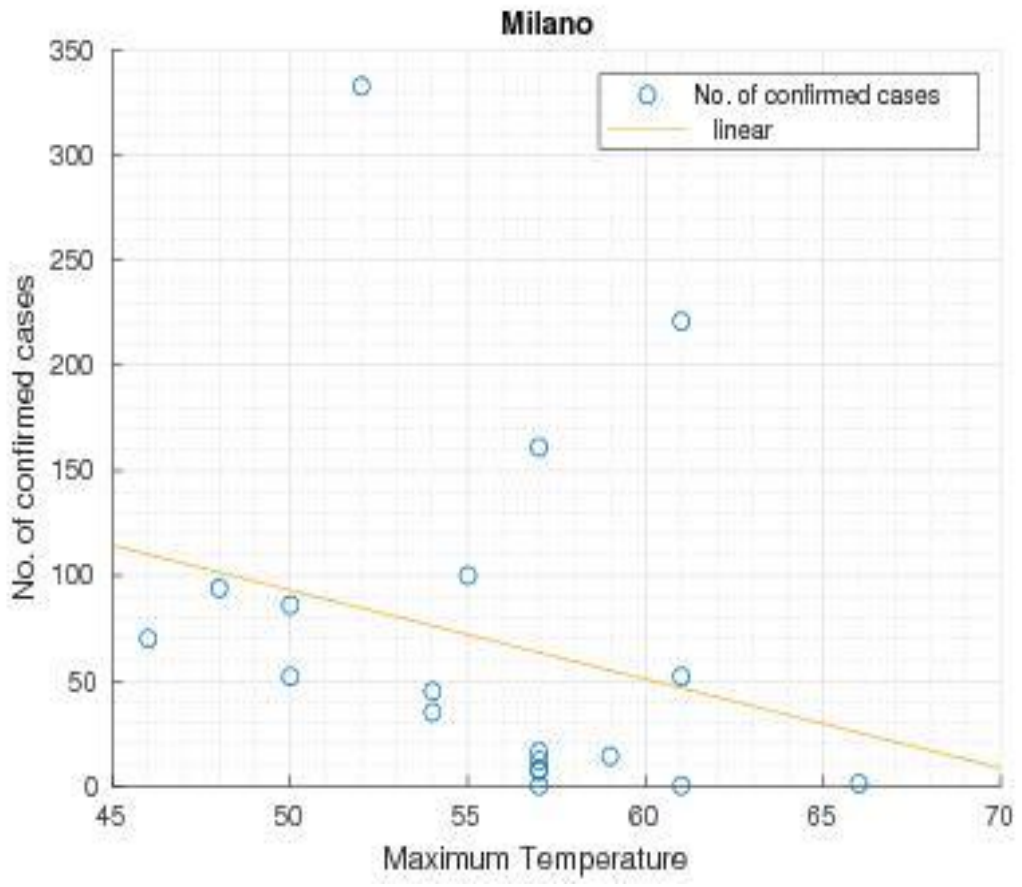

*Figure 18: Effect of maximum temperature in Milano*

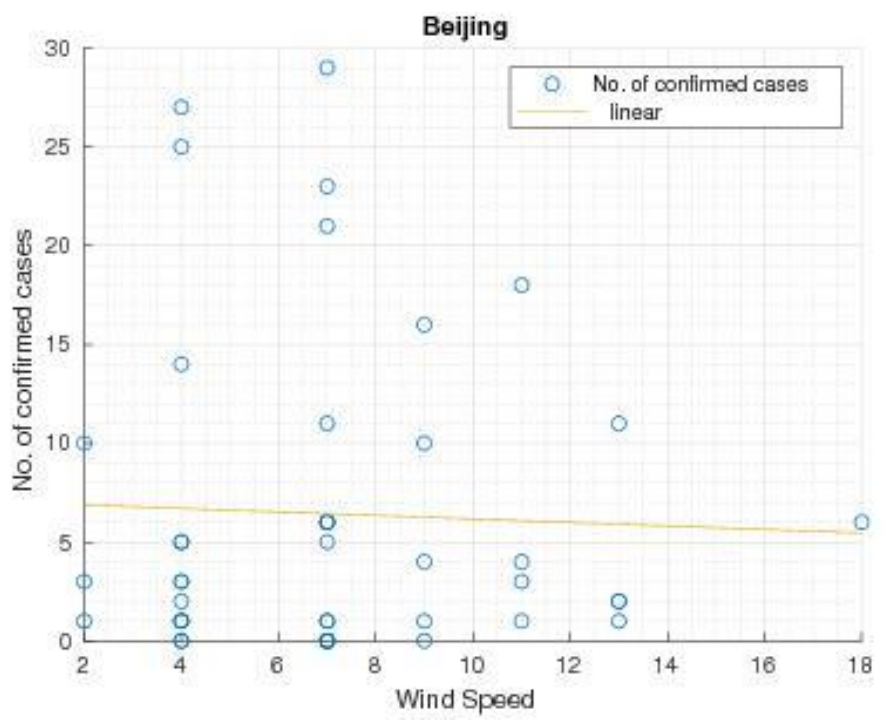

*Figure 19: Effect of maximum wind speed in Beijing*

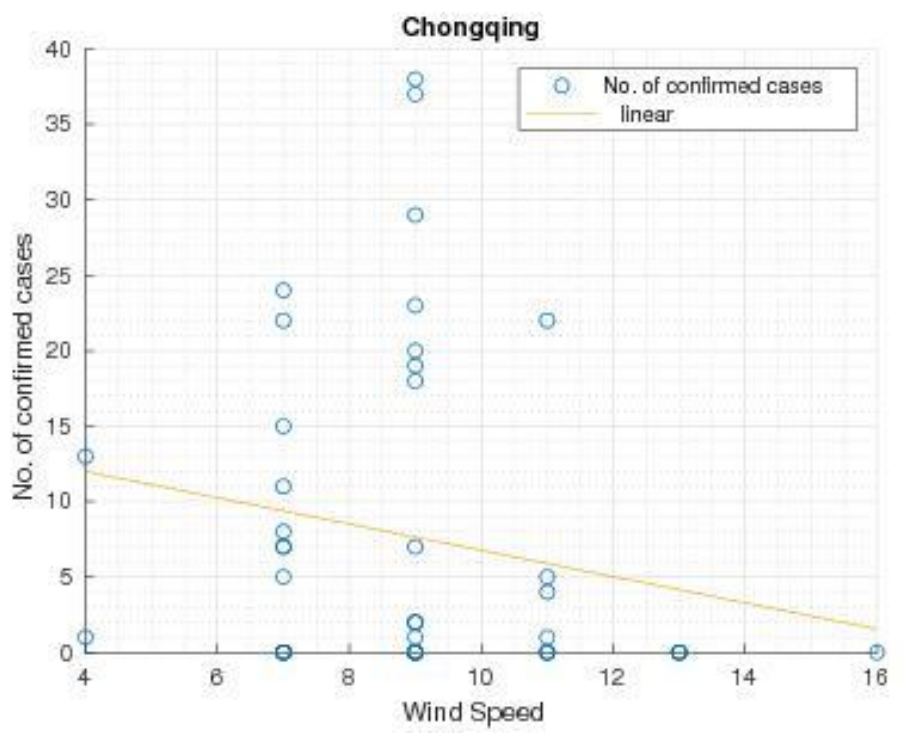

*Figure 20: Effect of maximum wind speed in Congqing*

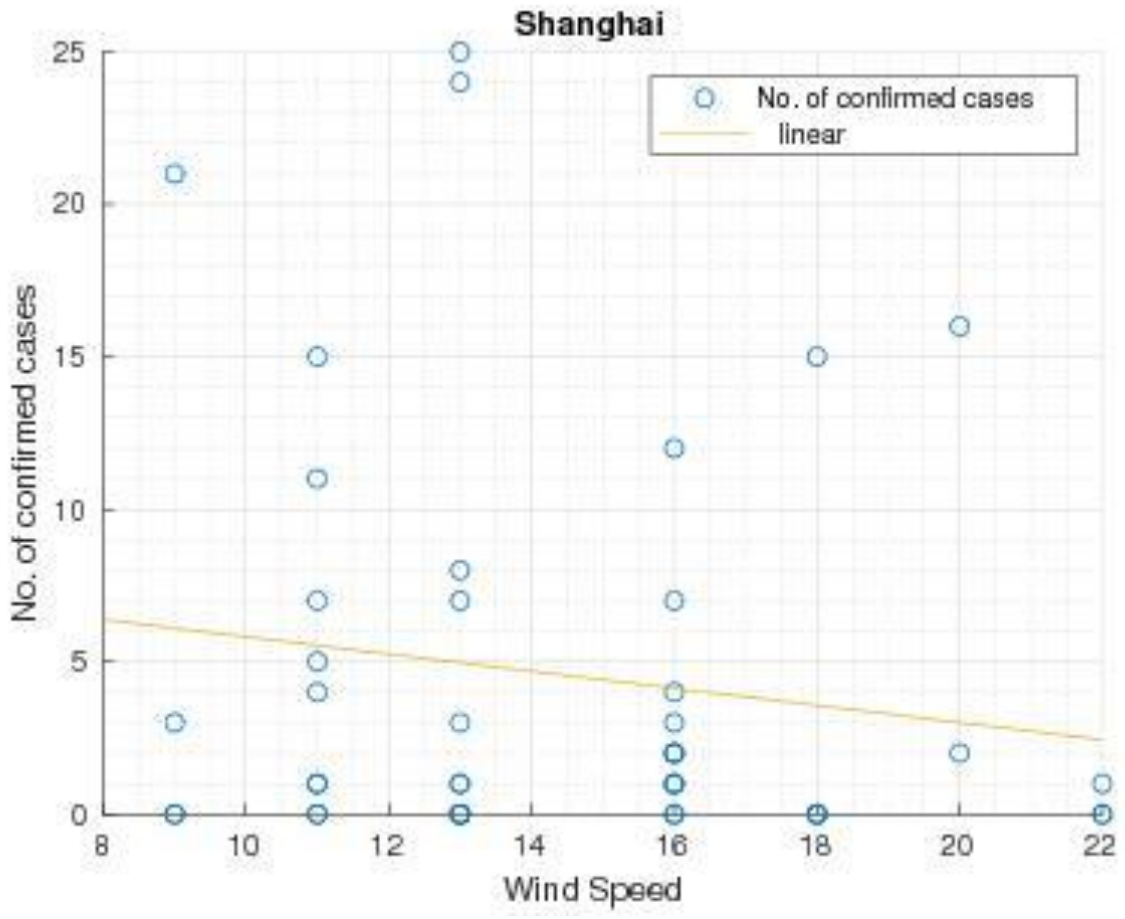

*Figure 21: Effect of maximum wind speed in Shanghai*

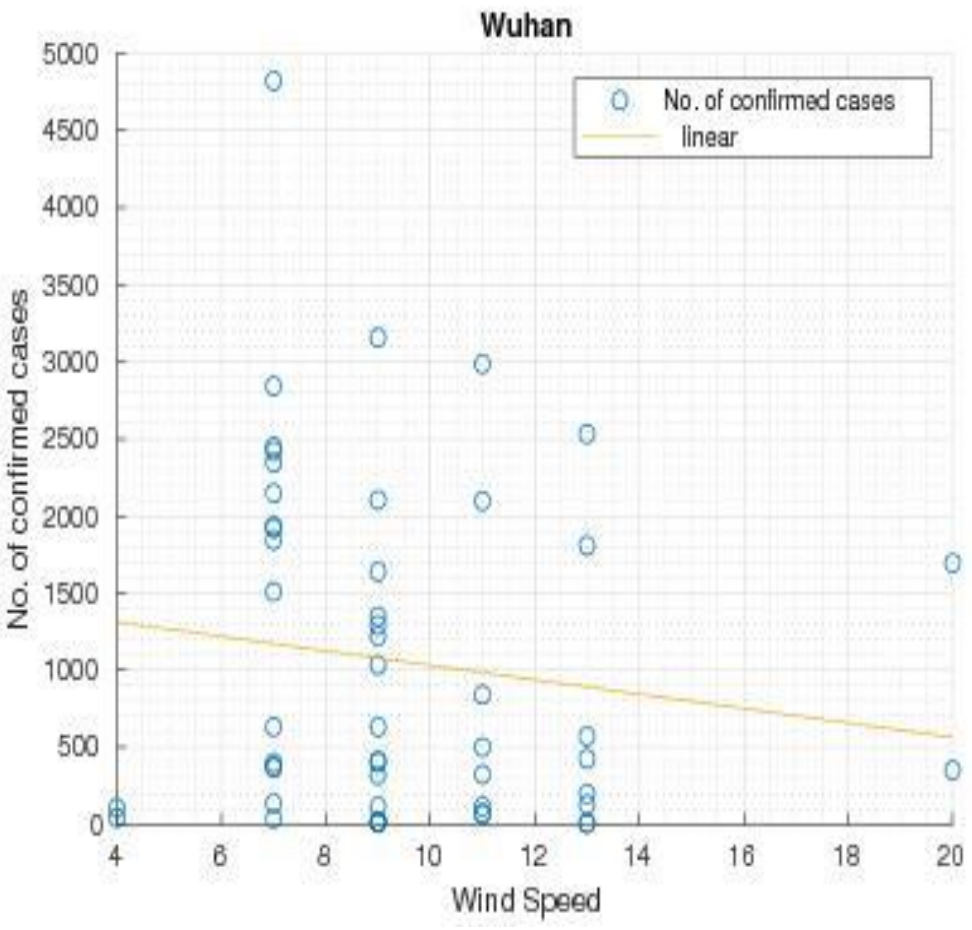

*Figure 22: Effect of maximum wind speed in Wuhan*

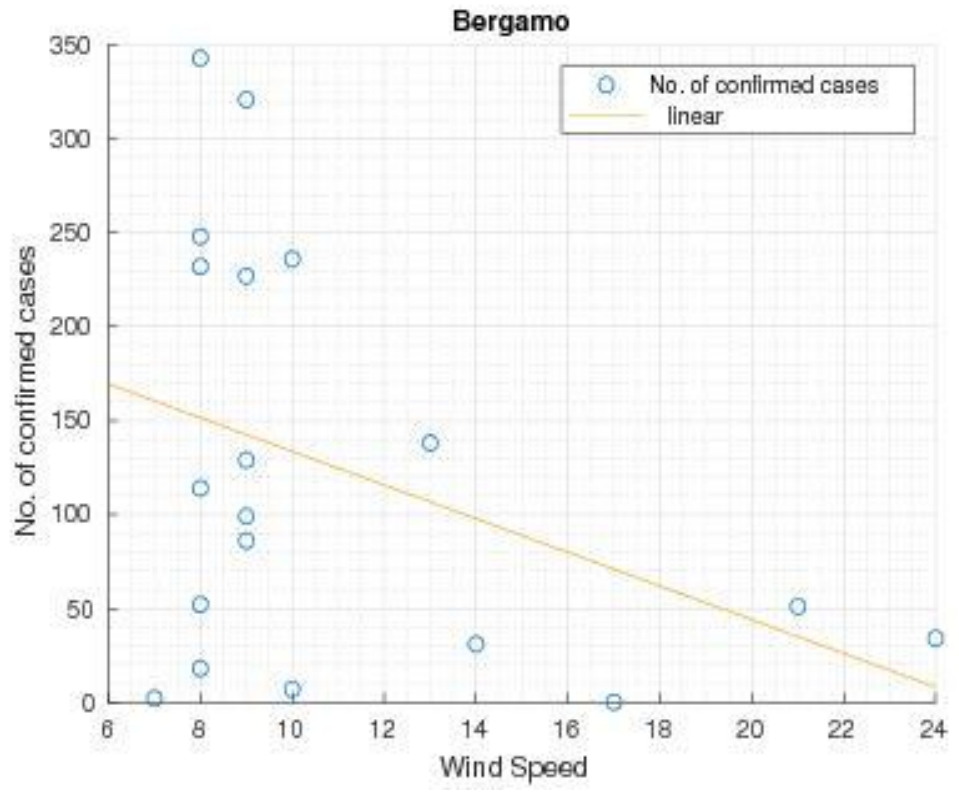

*Figure 23: Effect of maximum wind speed in Bergamo*

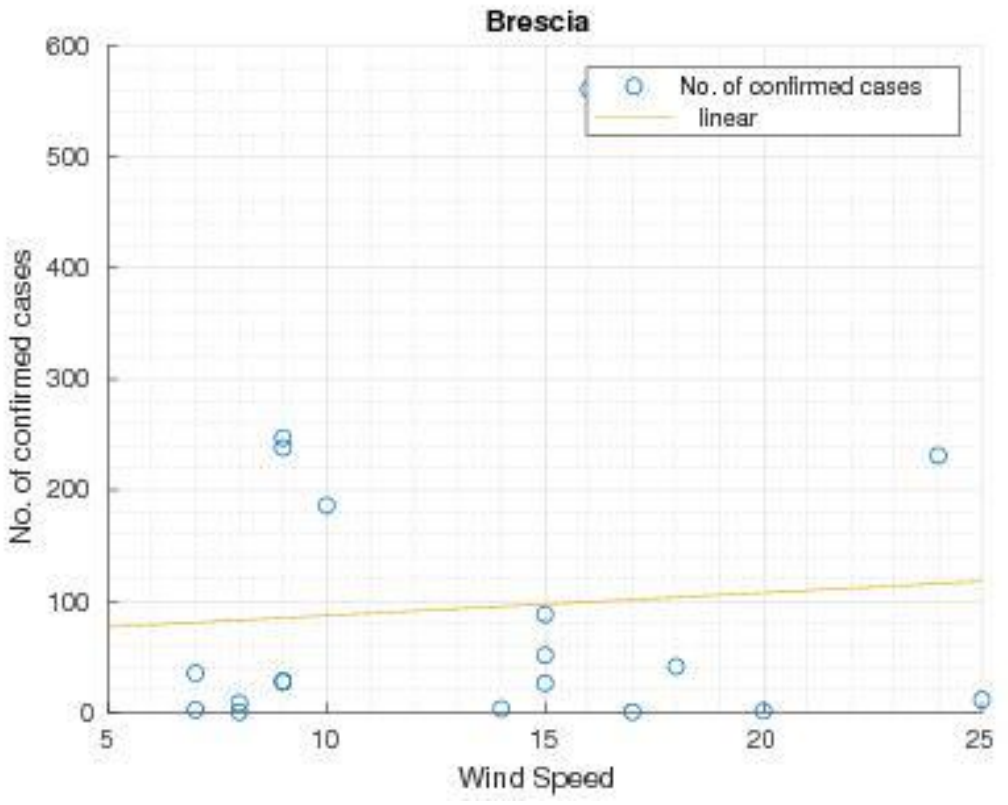

*Figure 24: Effect of maximum wind speed in Brescia*

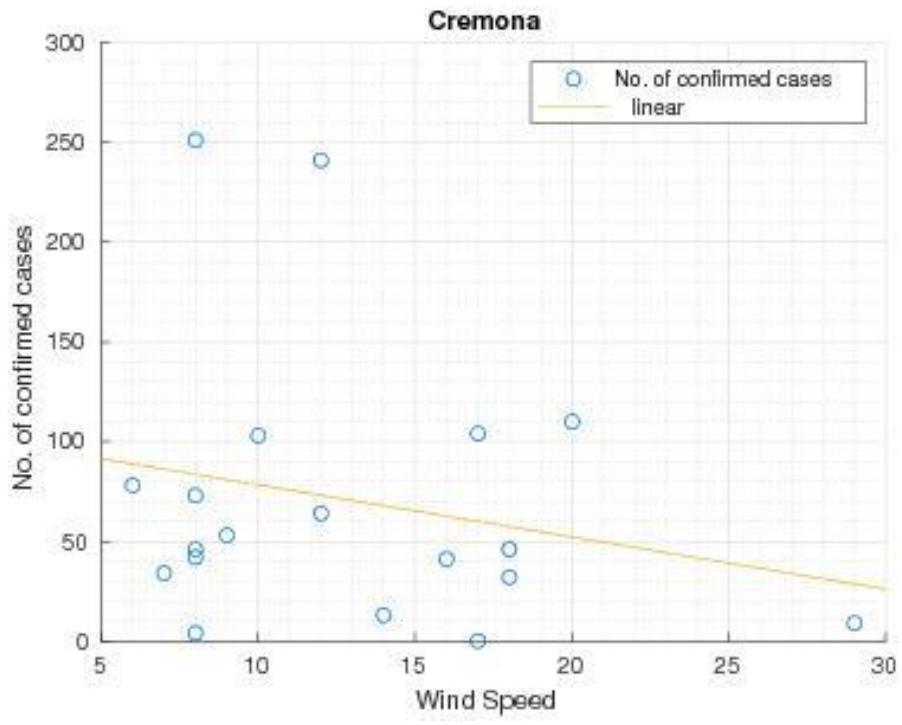

*Figure 25: Effect of maximum wind speed in Cremona*

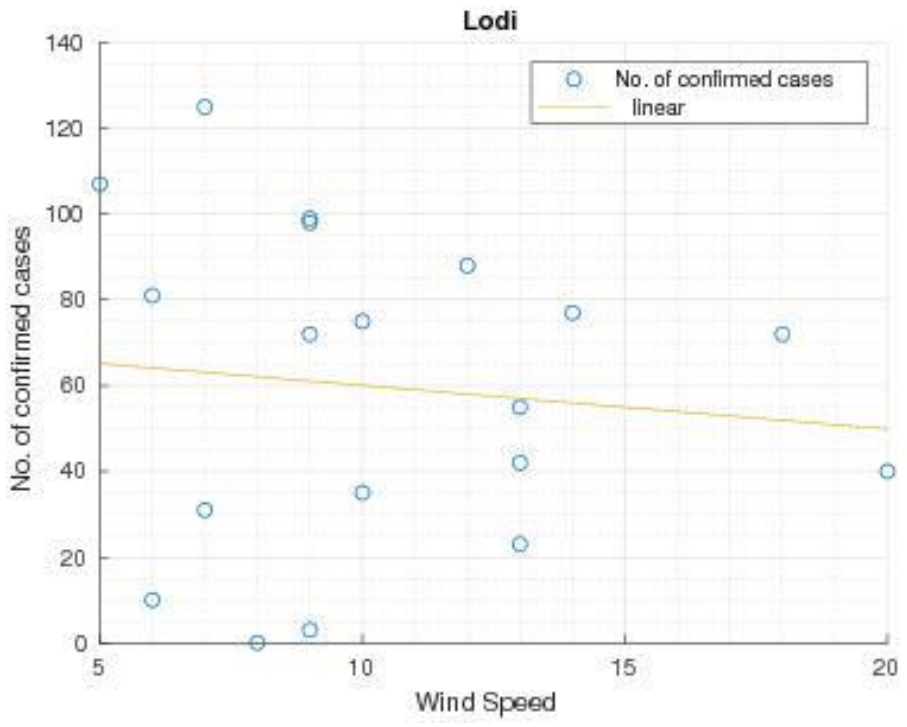

*Figure 26: Effect of maximum wind speed in Lodi*

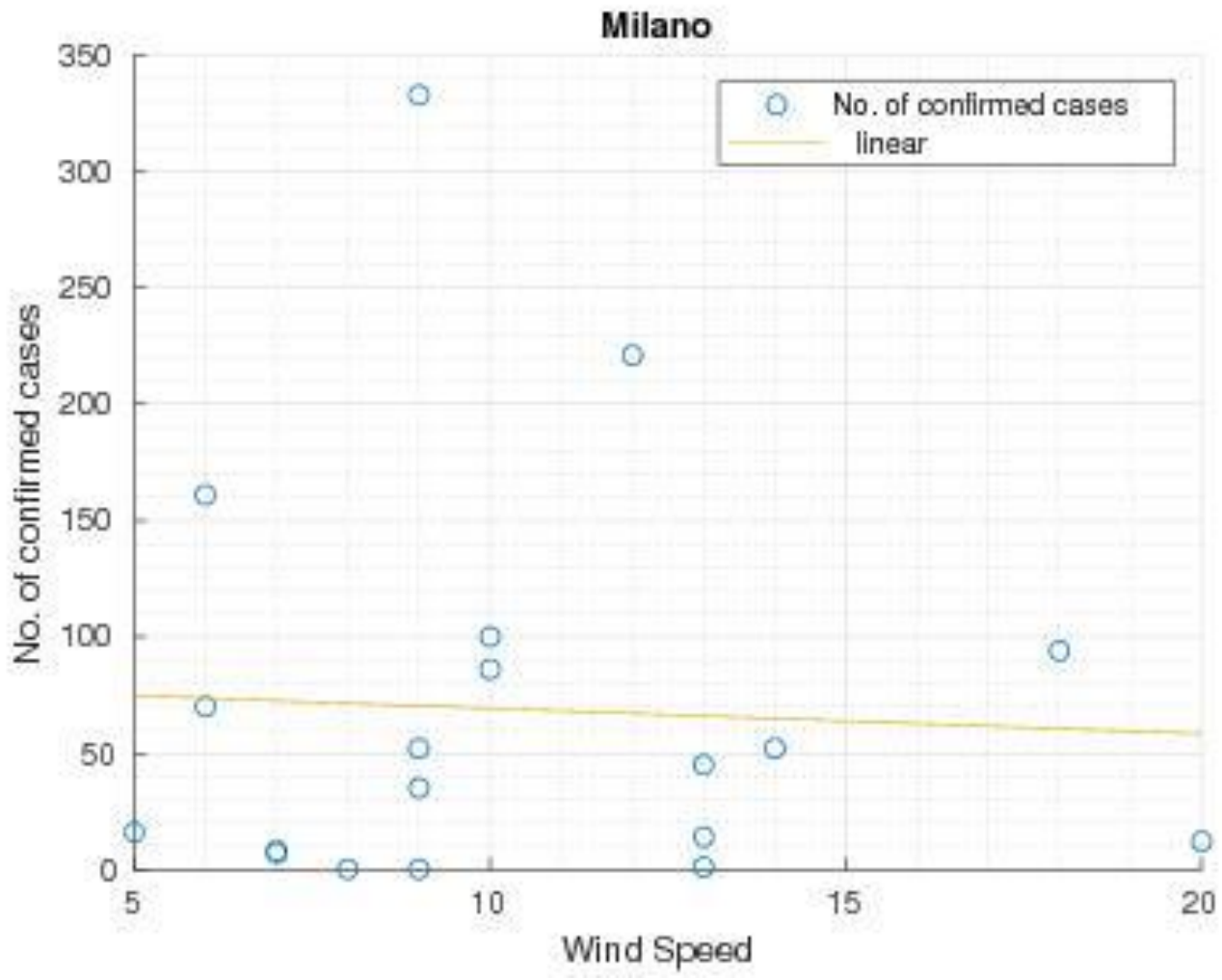

*Figure 27: Effect of maximum wind speed in Milano*

*Table 3: Values and interpretation of statistical indicators*

| Country | Place | Statistical Indicator | Highest RH on that day | | HIGHEST TEMP on that day (FAHRENHEIT) | | HIGHEST WIND SPEED on that day (mph) | |
|---|---|---|---|---|---|---|---|---|
| **China** | Beijing | r | 0.054 | Negligible | -0.2335 | Negligible | -0.0388 | Negligible |
| | | BF | 1.575587437 | Negligible | 1.429067674 | Negligible | 2.064372541 | Negligible |
| | Chongqing | r | 0.0344 | Negligible | -0.3925 | Low negative | -0.1919 | Negligible |
| | | BF | 2.290063003 | Negligible | 9.817365187 | Moderate | 1.129149822 | Negligible |
| | Shanghai | r | -0.0497 | Negligible | -0.325 | Low negative | -0.149 | Negligible |
| | | BF | 1.682343457 | Negligible | 3.547988527 | Moderate | 1.00598094 | Negligible |
| | Wuhan | r | 0.0282 | Negligible | 0.0421 | Negligible | -0.1338 | Negligible |
| | | BF | 2.55369917 | Negligible | 1.806152414 | Negligible | 1.002067883 | Negligible |
| Italy | Bergamo | r | 0.2925 | Negligible | -0.3515 | Low negative | -0.3854 | Low negative |
| | | BF | 1.097402009 | Negligible | 1.336500235 | Negligible | 1.569611454 | Negligible |
| | Brescia | r | 0.417 | Low negative | -0.2953 | Negligible | 0.0808 | Negligible |
| | | BF | 1.882892321 | Negligible | 1.105067996 | Negligible | 1.662529557 | Negligible |
| | Cremona | r | 0.1139 | Negligible | -0.1797 | Negligible | -0.2221 | Negligible |
| | | BF | 1.294281055 | Negligible | 1.030928263 | Negligible | 1.000186398 | Negligible |
| | Lodi | r | -0.3817 | Low negative | -0.2305 | Negligible | -0.1113 | Negligible |
| | | BF | 1.539157003 | Negligible | 1.002461915 | Negligible | 1.314347928 | Negligible |
| | Milano | r | 0.2569 | Negligible | -0.2436 | Negligible | -0.0504 | Negligible |
| | | BF | 1.025950716 | Negligible | 1.011070479 | Negligible | 2.478388031 | Negligible |

**Conclusion:** COVID-19 started in the month of December 2019, from China and is rapidly spreading to different countries of the world. Millions of people have already been infected by the virus SARS-CoV-2. Amidst the commotion, a belief is getting popular that the virus would die its own death with the arrival of summer season, but in the current paper, it has been found that the relationship between the effectiveness of virus and different environmental factors is not that strong. Hence, it can be concluded that the virus shows no sign as of now, to become dormant during summer days. The current piece of work is based on preliminary data that's available. A better relation can be predicted when more data become available.

**Acknowledgement:** Declarations of interest: none